\documentclass[a4paper,11pt]{article}
\pdfoutput=1 % if your are submitting a pdflatex (i.e. if you have
\usepackage{jheppub} % for details on the use of the package, please
\usepackage[T1]{fontenc} % if needed

\title{\boldmath Noether charge formalism for Weyl transverse gravity}

\author[a]{A. Alonso-Serrano}
\author[b,c]{L. J. Garay}
\author[a,d]{M. Li\v{s}ka,}

\affiliation[a]{Max-Planck-Institut f\"ur Gravitationsphysik (Albert-Einstein-Institut), \\Am M\"{u}hlenberg 1, 14476 Potsdam, Germany}
\affiliation[b]{Departamento de F\'{i}sica Te\'{o}rica and IPARCOS, Universidad Complutense de Madrid, 28040 Madrid, Spain}
\affiliation[c]{Instituto de Estructura de la Materia (IEM-CSIC), Serrano 121, 28006 Madrid, Spain}
\affiliation[d]{Institute of Theoretical Physics, Faculty of Mathematics and Physics, Charles University,
	\\V Hole\v{s}ovi\v{c}k\'{a}ch 2, 180 00 Prague 8, Czech Republic}

\emailAdd{ana.alonso.serrano@aei.mpg.de}
\emailAdd{luisj.garay@ucm.es}
\emailAdd{liska.mk@seznam.cz}

\abstract{Weyl transverse gravity is a gravitational theory that is invariant under transverse diffeomorphisms and Weyl transformations. It is characterised by having the same classical solutions as general relativity while solving some of its issues with the cosmological constant. In this work, we first find the Noether currents and charges corresponding to local symmetries of Weyl transverse gravity as well as a prescription for the symplectic form. We then employ these results to derive the first law of black hole mechanics in Weyl transverse gravity (both in vacuum and in the presence of a perfect fluid), identifying the total energy, the total angular momentum, and the Wald entropy of black holes. We further obtain the first law and Smarr formula for Schwarzschild-anti-de Sitter and pure de Sitter spacetimes, discussing the contributions of the varying cosmological constant, which naturally appear in Weyl transverse gravity. Lastly, we derive the first law of causal diamonds in vacuum.}

\begin{document}
\maketitle
\flushbottom

\section{Introduction}

The implementation of the Noether charge formalism in gravitational systems provides a systematic way to calculate conserved quantities in these theories~\cite{Wald:1990,Wald:1993,Wald:1994,Iyer:1996,Wald:2000}. Among the successes of this method we can count a formulation of the first law of black hole mechanics and a prescription for black hole entropy (known as Wald entropy) valid for all local, diffeomorphism (Diff) invariant theories of gravity~\cite{Wald:1993,Wald:1994,Iyer:1996}. It has also been employed to calculate the conserved charges corresponding to the Bondi-Metzner-Sachs symmetry group~\cite{Wald:2000}.

The seminal works on the gravitational Noether charge formalism were concerned only with local, Diff invariant theories. Here, we extend this analysis to gravity invariant under transverse diffeomorphisms and Weyl transformations (WTDiff). While there exist a large class of gravitational theories with a WTDiff symmetry group, we focus on the simplest one, Weyl transverse gravity (WTG).

The relevance of WTG lies in the fact that it represents a classical alternative to general relativity (GR), and even offers some advantages over it. The idea of WTG has first emerged from the observation that the construction of a consistent theory of self-interacting gravitons does not uniquely lead to GR. Instead, it also allows to arrive at a WTDiff invariant theory now known as WTG~\cite{Alvarez:2006,Barcelo:2014,Barcelo:2018}. Classical solutions of WTG and GR equations of motion are equivalent (although WTG in principle allows coupling to more general matter sources, as we discuss in subsection~\ref{matter}). However, WTG solves some of the problems connected with the cosmological constant, $\Lambda$, intrinsic to GR. Applying the formalism of the effective field theory to GR appears to yield unrealistically high estimate for $\Lambda$ coming from vacuum energy~\cite{Weinberg:1989,Polchinski:2006,Burgess:2013} (see, however,~\cite{Donoghue:2021} for an alternative viewpoint). Both WTG and unimodular gravity (a gauge fixed version of WTG) solve this problem, since vacuum energy does not gravitate in these theories~\cite{Barcelo:2014,Barcelo:2018}. Perhaps more worryingly, $\Lambda$ is radiatively unstable in GR~\cite{Padilla:2015,Carballo:2015}. In other words, even if one somehow fixes $\Lambda$ to be consistent with observations at the tree level, higher loop corrections will drastically alter its value. Hence, arriving at a $\Lambda$ value consistent with observations apparently requires infinite amount of fine-tuning. This challenges the rationale behind the effective field theory, i.e., that high energy physics does not significantly affect low energy observations. However, Weyl invariance of WTG ensures radiative stability of $\Lambda$~\cite{Carballo:2015}, avoiding the problems with validity of the effective field theory approach.

Within this framework, developing a Noether charge formalism for WTG provides, on one side, a consistency check for this theory. On the other side, it offers a practical tool to calculate conserved quantities in WTG, aiding its further development.

Another interest of this analysis comes from the results of applying the Noether charge formalism to the case of causal diamonds. One then obtains a first law of causal diamond mechanics and even an expression for their Wald entropy, both analogous to the results known for black holes~\cite{Jacobson:2019}. Conversely, considering Wald entropy and Unruh temperature associated with local causal diamonds allows one to derive equations governing gravitational dynamics~\cite{Jacobson:2015,Bueno:2017,Svesko:2019}. However, the resulting dynamics is consistent with unimodular gravity, which is invariant only under transverse diffeomorphisms~\cite{Alonso:2020a,Alonso:2021}. Furthermore, gravitational dynamics derived from thermodynamics is invariant under Weyl transformations~\cite{preparation} and, thus, has exactly the same symmetries as WTG. In this way, starting with Wald entropy obtained from the Diff invariant Lagrangian of GR one arrives at equations of gravitational dynamics consistent with WTDiff invariant WTG. This motivates us to see whether an expression for Wald entropy can be recovered even for WTG.

In this work, we develop a new version of the Noether charge formalism adapted to WTG. We start by reviewing the main features of WTDiff invariant theories in section~\ref{WTG}. Then, in section~\ref{Noether}, we derive expressions for the Noether current and charge. In section~\ref{physics}, we employ these expressions to derive the first law of black hole mechanics and an expression for Wald entropy. Furthermore, we analyse the cases of Schwarzschild-anti-de Sitter spacetime and de Sitter spacetime which highlight some differences between Diff and WTDiff invariant theories. Section~\ref{sourced} concerns the first law of black hole mechanics in the presence of matter fields. In section~\ref{conformal}, we apply our formalism to conformal Killing vectors and derive the first law of causal diamonds. Finally, section~\ref{discussion} sums up our findings and discusses possible further applications of the formalism.

Unless otherwise specified, we work in an arbitrary spacetime dimension $n$ and use metric signature $(-,+,...,+)$. We set $G=c=\hbar=k_B=1$. Definitions of the curvature-related quantities follow~\cite{MTW}.

\section{Weyl transverse gravity}
\label{WTG}

In this section, we first review the basics of WTG and definitions of some auxiliary quantities useful for working in a WTDiff invariant setting. Then, we turn to the coupling of WTG to matter fields, highlighting the differences compared to Diff invariant theories.

\subsection{Vacuum theory}

The action for vacuum WTG in $n$ spacetime dimensions reads~\cite{Alvarez:2013a,Oda:2017}
\begin{align}
\nonumber S_{\text{WTG}}=&\frac{1}{16\pi}\int_{V}\left[R+\frac{\left(n-1\right)\left(n-2\right)}{n^2}g^{\alpha\beta}\partial_{\alpha}\ln\frac{\sqrt{-\mathfrak{g}}}{\omega}\partial_{\beta}\ln\frac{\sqrt{-\mathfrak{g}}}{\omega}\right]\left(\frac{\sqrt{-\mathfrak{g}}}{\omega}\right)^{\frac{2}{n}}\omega\text{d}^nx \\
&-\frac{1}{8\pi}\int_{V}\lambda\omega\text{d}^nx+\frac{1}{8\pi}\frac{n-1}{n}\oint_{\partial V}\left(\frac{\sqrt{-\mathfrak{g}}}{\omega}\right)^{\frac{1}{n}}\partial_{\mu}\ln\frac{\sqrt{-\mathfrak{g}}}{\omega}n^{\mu}\omega\text{d}^{n-1}x, \label{action}
\end{align}
where $\mathfrak{g}$ is the determinant of the metric, $V$ is the domain of integration, $\partial V$ denotes its boundary with a unit normal $n^{\mu}$, and we introduced a non-dynamical volume $n$-form \mbox{$\boldsymbol{\omega}=\omega\left(x\right)\text{d}x^{0}\wedge\text{d}x^{1}\wedge...\wedge\text{d}x^{n-1}$}, with $\omega\left(x\right)$ being a strictly positive function. The presence of such a privileged background $n$-form is necessary for any WTDiff invariant theory of gravity~\cite{Barcelo:2018} (unless one chooses to introduce dynamical degrees of freedom besides the metric~\cite{Oda:2017,Jirousek:2019}). In principle, it is possible to introduce dynamics for the volume $n$-form $\boldsymbol{\omega}$, without affecting the gravitational dynamics of WTDiff invariant gravity or the radiative stability of $\Lambda$~\cite{Carballo:2020}. However, for our current purposes the underlying dynamics of the background volume $n$-form plays no role and we will treat it as non-dynamical.

The second integral in the action contains an arbitrary constant $\lambda$ and, as it is completely independent of the metric (the only dynamical quantity in vacuum), it does not contribute to gravitational equations of motion and does not even enter higher order effective field theory calculations. Therefore, unless specified otherwise, we simply set $\lambda=0$ in the following without any loss of generality.

To rewrite the action for WTG in a simpler form, we introduce a WTDiff invariant (but not Diff invariant) auxiliary metric
\begin{equation}
\tilde{g}_{\mu\nu}=\left(\sqrt{-\mathfrak{g}}/\omega\right)^{-2/n}g_{\mu\nu}.
\end{equation}
One may notice that $\tilde{g}_{\mu\nu}$ is just the dynamical metric $g_{\mu\nu}$ restricted to the unimodular gauge, $\sqrt{-\mathfrak{g}}=\omega$. In principle, it is possible to formulate WTG in the unimodular gauge and consider $\tilde{g}_{\mu\nu}$ as the dynamical variable. However, we impose no such gauge restriction and only use $\tilde{g}_{\mu\nu}$ as a convenient notational device. We further note that raising and lowering of indices is always done with the dynamical metric, $g_{\mu\nu}$ (the obvious exception being $\tilde{g}^{\mu\nu}$ which we define as an inverse of $\tilde{g}_{\mu\nu}$).

The Weyl connection corresponding to $\tilde{g}_{\mu\nu}$ reads
\begin{equation}
\tilde{\Gamma}^{\mu}_{\;\:\nu\rho}=\Gamma^{\mu}_{\;\:\nu\rho}-\frac{1}{n}\left(\delta^{\mu}_{\nu}\delta^{\alpha}_{\rho}+\delta^{\mu}_{\rho}\delta^{\alpha}_{\nu}-g_{\nu\rho}g^{\mu\alpha}\right)\partial_{\alpha}\ln\frac{\sqrt{-\mathfrak{g}}}{\omega}.
\end{equation}
Using $\tilde{g}_{\mu\nu}$ and $\tilde{\Gamma}^{\mu}_{\;\:\nu\rho}$ we can write the action of WTG in a simpler form,
\begin{equation}
\label{WTG action}
S_{\text{WTG}}=\frac{1}{16\pi}\int_{V}\tilde{R}\omega\text{d}^{n}x,
\end{equation}	
where $\tilde{R}=\tilde{g}^{\mu\nu}\tilde{R}_{\mu\nu}$ and $\tilde{R}_{\mu\nu}$ is the Ricci curvature tensor constructed from the auxiliary (WTDiff invariant) Riemann tensor
\begin{align}
\nonumber \tilde{R}^{\mu}_{\;\:\nu\rho\sigma}=& 2\tilde{\Gamma}^{\mu}_{\;\:\nu[\sigma,\rho]}+2\tilde{\Gamma}^{\mu}_{\;\:\lambda[\rho}\tilde{\Gamma}^{\lambda}_{\;\:\sigma]\nu}\\
\nonumber =&R^{\mu}_{\;\:\nu\rho\sigma}+\frac{2}{n}\left(\delta^{\alpha}_{\nu}\delta^{\mu}_{[\rho}\delta^{\beta}_{\sigma]}+g^{\mu\alpha}g_{\nu[\sigma}\delta^{\beta}_{\rho]}\right)\partial_{\alpha}\partial_{\beta}\ln\frac{\sqrt{-\mathfrak{g}}}{\omega} \\
&+\frac{2}{n^2}\left(g^{\alpha\beta}g_{\nu[\rho}\delta^{\mu}_{\sigma]}+\delta^{\alpha}_{\nu}\delta^{\mu}_{[\rho}\delta^{\beta}_{\sigma]}+g^{\mu\alpha}\delta^{\beta}_{[\rho}g_{\sigma]\nu}\right)\partial_{\alpha}\ln\frac{\sqrt{-\mathfrak{g}}}{\omega}\partial_{\beta}\ln\frac{\sqrt{-\mathfrak{g}}}{\omega}.
\end{align}
The WTG action is invariant with respect to Weyl transformations
\begin{equation}
\delta g_{\mu\nu}=e^{2\sigma}g_{\mu\nu},
\end{equation}
where $\sigma$ is an arbitrary spacetime function. Likewise, it is invariant with respect to transverse diffeomorphisms. However, the usual transversality condition, $\nabla_{\mu}\xi^{\mu}=0$, where $\xi^{\mu}$ denotes the diffeomorphism generator, is not Weyl invariant. Thus, it cannot be satisfied in every Weyl frame simultaneously, making it unsuitable for WTG. Instead, one must define transversality with respect to the Weyl invariant covariant derivative. Transverse diffeomorphisms then obey
\begin{align}
\delta g_{\mu\nu}=&2\nabla_{(\nu}\xi_{\mu)}, \\
\tilde{\nabla}_{\mu}\xi^{\mu}=&0\qquad\iff\qquad\nabla_{\mu}\xi^{\mu}=\xi^{\mu}\partial_{\mu}\ln\frac{\sqrt{-\mathfrak{g}}}{\omega}.
\end{align}
The theory is not invariant under longitudinal diffeomorphisms (with $\tilde{\nabla}_{\mu}\xi^{\mu}\ne0$).

Varying the action with respect to $g^{\mu\nu}$ yields the vacuum equations of motion of WTG
\begin{equation}
\label{vacuum EoMs}
\tilde{R}_{\mu\nu}-\frac{1}{n}\tilde{R}\tilde{g}_{\mu\nu}=0,
\end{equation}
which are traceless. We can restate these equations of motion in a form more reminiscent of the standard Einstein equations. Invoking the contracted Bianchi identities,
\begin{equation}
\label{Bianchi}
2\tilde{g}^{\nu\rho}\tilde{\nabla}_{\nu}\tilde{R}_{\mu\rho}=\tilde{\nabla}_{\mu}\tilde{R},
\end{equation}
we find that on shell
\begin{equation}
\left(n-2\right)/\left(2n\right)\tilde{R}\tilde{g}_{\mu\nu}=\Lambda\tilde{g}_{\mu\nu},
\end{equation}
where $\Lambda$ is an arbitrary integration constant. Finally, subtracting the previous equality from the traceless equations of motion yields
\begin{equation}
\tilde{R}_{\mu\nu}-\frac{1}{2}\tilde{R}\tilde{g}_{\mu\nu}+\Lambda\tilde{g}_{\mu\nu}=0.
\end{equation}
By comparison with the Einstein equations, we can see that $\Lambda$ plays the role of the cosmological constant. In contrast with GR, $\Lambda$ has no connection with the constant $\lambda$ present in the Lagrangian, which we already set to $0$. Furthermore, $\Lambda$ is naturally allowed to vary between solutions of the equations of motion, since it is an integration constant.

\subsection{Coupling to matter fields}
\label{matter}

Let us now discuss the behaviour of matter fields coupled to WTG. We start by introducing a WTDiff invariant matter action and the form of WTG equations of motion with matter. Lastly, we discuss local energy-momentum conservation.

Consider a matter field minimally coupled to WTG. The corresponding WTDiff invariant matter action takes form
\begin{equation}
S_{\psi}=\int_{V}\left(\sqrt{-\mathfrak{g}}/\omega\right)^{2k/n}L_{\psi}\omega\text{d}^{n}x,
\end{equation}
where $L_{\psi}$ is a function of the matter variables $\psi$, their Weyl invariant covariant derivatives, and $k$ contravariant metric tensors, $g^{\mu\nu}$. The factor $\left(\sqrt{-\mathfrak{g}}/\omega\right)^{2k/n}$ then ensures the overall Weyl invariance of the action, compensating the behaviour of  $g^{\mu\nu}$ under Weyl transformations\footnote{We might instead just replace $g^{\mu\nu}$ by $\tilde{g}^{\mu\nu}$ in a Diff invariant matter Lagrangian. However, since we work with the standard definition of the energy-momentum tensor, the notation we chose is more practical for our purposes.}. By definition, the Weyl transformations affect only the gravitational sector of the theory and matter variables are invariant under them. If more than one matter field is present, $S_{\psi}$ is just a sum of several terms of this form (including possible interaction terms).

By simultaneously varying the WTG and matter actions with respect to $g^{\mu\nu}$ we obtain the full gravitational equations of motion
\begin{equation}
\label{EoMs}
\tilde{R}_{\mu\nu}-\frac{1}{n}\tilde{R}\tilde{g}_{\mu\nu}=8\pi\left(\frac{\sqrt{-\mathfrak{g}}}{\omega}\right)^{2\frac{k-1}{n}}\left(T_{\mu\nu}-\frac{1}{n}Tg_{\mu\nu}\right),
\end{equation}
where we define the energy-momentum tensor $T_{\mu\nu}$ by the standard Hilbert prescription\footnote{One might instead consider an energy-momentum tensor defined with respect to the auxiliary metric~\cite{Carballo:2020}
\begin{equation}
\tilde{T}_{\mu\nu}=-2\frac{\partial\left[\left(\sqrt{-\mathfrak{g}}/\omega\right)^{2k/n}L_{\psi}\right]}{\partial\tilde{g}^{\mu\nu}}+\left(\sqrt{-\mathfrak{g}}/\omega\right)^{2k/n}L_{\psi}\tilde{g}_{\mu\nu},
\end{equation}
which has the advantage of being Weyl invariant. However, we choose the standard definition as it involves variations with respect to the full dynamical metric.}
\begin{equation}
T_{\mu\nu}=-2\frac{\partial L_{\psi}}{\partial g^{\mu\nu}}+L_{\psi}g_{\mu\nu},
\end{equation}
and $T=T_{\mu\nu}g^{\mu\nu}$ denotes its trace. From the Hilbert prescription and Weyl invariance of matter variables, we can easily see the behaviour of the energy-momentum tensor under Weyl transformations
\begin{equation}
T'_{\mu\nu}=e^{-2\frac{k-1}{n}\sigma}T_{\mu\nu}.
\end{equation}
Thus, the right hand side of the WTG equations of motion~\eqref{EoMs} is Weyl invariant.

Let us now briefly address the question of local energy-momentum conservation in WTDiff invariant gravity. Since the matter and gravitational actions are WTDiff invariant rather than Diff invariant, the energy-momentum tensor is not guaranteed to be divergence-free by the symmetries of the action (nor, conversely, by the combination of the equations of motion and the contracted Bianchi identities). Nevertheless, invariance with respect to transverse diffeomorphisms still ensures~\cite{Alvarez:2013a}
\begin{equation}
\label{div T}
8\pi\tilde{\nabla}_{\nu}\left[\left(\sqrt{-\mathfrak{g}}/\omega\right)^{2k/n}T_{\mu}^{\;\:\nu}\right]=\tilde{\nabla}_{\mu}\mathcal{J},
\end{equation}
for some scalar function $\mathcal{J}$. Clearly, nonzero $\mathcal{J}$ implies that energy-momentum is not locally conserved. The condition $\tilde{\nabla}_{\nu}T_{\mu}^{\;\:\nu}=0$ is anyway fulfilled for most of the often considered matter sources, thanks to the matter equations of motion, but this is not a requirement of the theory. The freedom to break local energy-momentum conservation by introducing $\mathcal{J}\ne0$ has been exploited, e.g. to propose a mechanism for cosmic acceleration~\cite{Josset:2017,Perez:2018}. In the present work, we therefore assume the most general situation, $\mathcal{J}\ne0$.

\section{The Noether charge formalism}
\label{Noether}

Upon reviewing the properties of WTG, we proceed to develop the Noether charge formalism applicable to it. Our goal is to identify conserved quantities corresponding to symmetries of a given spacetime. Before specialising to WTG, we briefly review general aspects of the Noether charge formalism. Then, we consider general variations of the vacuum WTG action and calculate the symplectic potential and the symplectic current. Next, we derive the Noether currents and charges corresponding to symmetry transformations, i.e., Weyl transformations and transverse diffeomorphisms. Lastly, we find a formula for a general variation of a Hamiltonian corresponding to evolution along a transverse diffeomorphism generator.

\subsection{General formalism}
\label{general}

We start by briefly presenting the Noether charge formalism in a general setting in order to establish the notation and basic concepts. On a manifold equipped with some volume form, $\boldsymbol{\varepsilon}$, define a Lagrangian $L$, constructed from a collection of dynamical variables, $\phi$, some non-dynamical variables, $\gamma$, and their covariant derivatives, $\nabla_{\mu}$, which obeys $\nabla_{\mu}\boldsymbol{\varepsilon}=0$ (clearly, $\nabla_{\mu}$ is not unique). The change of $L$ under an arbitrary variation of the dynamical variables, $\delta_1 \phi$, equals
\begin{equation}
\label{dL g}
\delta_1 L=A_{\phi}\delta_1 \phi+\nabla_{\mu}\theta^{\mu}\left[\delta_{1}\right],
\end{equation}
where $A_{\phi}=0$ are the equations of motion, and, by virtue of the Gauss theorem, $\nabla_{\mu}\theta^{\mu}\left[\delta_{1}\right]$ contributes only a boundary integral to the variation of the action. We call $\theta^{\mu}$ the symplectic potential~\cite{Wald:1990}.

Now perform a second arbitrary variation of the dynamical variables, $\delta_2\phi$, independent of $\delta_1\phi$. The commutator of the variations applied to the Lagrangian reads
\begin{equation}
\left(\delta_1\delta_2-\delta_2\delta_1\right)L=\delta_1A_{\phi}\delta_2 \phi-\delta_2A_{\phi}\delta_1 \phi+\nabla_{\mu}\Omega^{\mu}\left[\delta_1,\delta_2\right],
\end{equation}
where
\begin{equation}
\label{omega g}
\Omega^{\mu}\left[\delta_{1},\delta_{2}\right]=\delta_{1}\theta^{\mu}\left[\delta_{2}\right]-\delta_{2}\theta^{\mu}\left[\delta_{1}\right],
\end{equation}
is known as the symplectic current~\cite{Wald:1990}. The integral of $\Omega^{\mu}\left[\delta_{1},\delta_{2}\right]$ over an initial data surface $\mathcal{C}$ then yields a symplectic form\footnote{Technically, the form defined in this way can be degenerate~\cite{Wald:1990}. To obtain a true symplectic form, one must restrict it from the space of field configurations to the phase space. However, this subtlety is not important for our purposes.}
\begin{equation}
\Omega\left[\delta_{1},\delta_{2}\right]=\int_{\mathcal{C}}\Omega^{\mu}\left[\delta_{1},\delta_{2}\right]\text{d}\mathcal{C}_{\mu}.
\end{equation}
If one of the variations is generated by a vector field $\xi^{\mu}$, i.e., $\delta_1\phi=\pounds_{\xi}\phi$, and there exists a Hamiltonian $H_{\xi}$ corresponding to evolution along $\xi^{\mu}$, the Hamilton equations of motion give us the variation of $H_{\xi}$ for an arbitrary variation, $\delta_{2}\phi=\delta\phi$, as
\begin{equation}
\label{dH0}
\delta H_{\xi}=\Omega\left[\pounds_{\xi},\delta\right].
\end{equation}

Now consider a variation $\hat{\delta}\phi$ corresponding to a local symmetry of the Lagrangian $L$. The change of $L$ under such a variation is a total divergence, $\hat{\delta}L=\nabla_{\mu}\alpha^{\mu}\big[\hat{\delta}\big]$, for some vector $\alpha^{\mu}$. The Noether current corresponding to a local symmetry reads~\cite{Wald:1990}
\begin{equation}
\label{j}
j^{\mu}\big[\hat{\delta}\big]=\theta^{\mu}\big[\hat{\delta}\big]-\alpha^{\mu}\big[\hat{\delta}\big].
\end{equation}
From now on we drop the argument $\hat{\delta}$ unless it is needed. The covariant divergence of $j^{\mu}$ is proportional to the equations of motion (see~\eqref{dL g})
\begin{equation}
\nabla_{\mu}j^{\mu}=\nabla_{\mu}\theta^{\mu}-\nabla_{\mu}\alpha^{\mu}=\nabla_{\mu}\theta^{\mu}-\hat{\delta}L=-A_{\phi}\hat{\delta}\phi,
\end{equation}
and, thus, vanishes on shell. The Noether charge corresponding to $j^{\mu}$ is just its integral over an initial data surface $\mathcal{C}$~\cite{Wald:1990}
\begin{equation}
Q=\int_{\mathcal{C}}j^{\mu}\text{d}\mathcal{C}_{\mu}.
\end{equation}

In the following subsections, we apply this general formalism to the case of WTG.

\subsection{WTG symplectic potential and form}

\paragraph{Symplectic potential.}

We begin by computing the symplectic potential, which directly leads to expressions for both the Noether current and the symplectic form. Our starting point is a general variation of the WTG Lagrangian~\eqref{WTG action} with respect to the dynamical metric, $g_{\mu\nu}$,
\begin{equation}
\label{dL}
\delta L=-\frac{1}{16\pi}\left[\left(\sqrt{-\mathfrak{g}}/\omega\right)^{2/n}\tilde{R}^{\mu\nu}-\frac{1}{n}\left(\sqrt{-\mathfrak{g}}/\omega\right)^{-2/n}\tilde{R}\tilde{g}^{\mu\nu}\right]\delta g_{\mu\nu}+\tilde{\nabla}_{\mu}\theta^{\mu}.
\end{equation}
The first term is proportional to the WTG equations of motion and vanishes on shell. The second one is a Weyl invariant divergence that contributes only a surface term to the variation of the action and, thus, does not affect the equations of motion\footnote{Let us briefly clarify the use of the Gauss theorem in this case. For an integral of a Weyl covariant divergence of any Weyl invariant vector $W^{\mu}$, we find
\begin{equation}
\int_{\Omega}\tilde{\nabla}_{\mu}W^{\mu}\omega_{\alpha_1\dots\alpha_n}=\int_{\Omega}\left(\omega\partial_{\mu}W^{\mu}+W^{\mu}\partial_{\mu}\omega\right)\epsilon_{\alpha_1\dots\alpha_n}=\int_{\Omega}\partial_{\mu}\left(\omega W^{\mu}\right)\epsilon_{\alpha_1\dots\alpha_n},
\end{equation}
where $\epsilon_{\alpha_1\dots\alpha_n}$ is the $n$-dimensional antisymmetrisation symbol and $\omega_{\alpha_1\dots\alpha_n}=\omega\epsilon_{\alpha_1\dots\alpha_n}$ is the background volume element. The Gauss theorem implies
\begin{equation}
\int_{\Omega}\partial_{\mu}\left(\omega W^{\mu}\right)\text{d}^{n}x=\int_{\partial\Omega}W^{\mu}n_{\mu}n^{\alpha_1}\omega_{\alpha_1\dots\alpha_n},
\end{equation}
where $n^{\mu}$ denotes a unit normal to $\partial\Omega$. As a normal vector, $n^{\mu}$ transforms as $n'^{\mu}=e^{-\sigma}n^{\mu}$ under Weyl transformations. If $n^{\mu}$ is a coordinate vector, we can write
\begin{equation}
\int_{\partial\Omega}W^{\mu}n_{\mu}n^{\alpha_1}\omega_{\alpha_1\dots\alpha_n}=\int_{\partial\Omega}\left(\sqrt{-\mathfrak{g}}/\omega\right)^{-1/n}W^{\mu}n_{\mu}\text{d}^{n-1}x.
\end{equation}}. 
The symplectic potential $\theta^{\mu}$ reads
\begin{equation}
\label{theta}
\theta^{\mu}\left[\delta\right]=\frac{1}{16\pi}\left(\frac{\sqrt{-\mathfrak{g}}}{\omega}\right)^{\frac{4}{n}}\left(g^{\mu\nu}g^{\rho\sigma}-g^{\mu\sigma}g^{\nu\rho}\right)\tilde{\nabla}_{\sigma}\delta\tilde{g}_{\nu\rho}.
\end{equation}
The variation of the auxiliary metric $\delta\tilde{g}_{\nu\rho}$ equals
\begin{equation}
\delta\tilde{g}_{\nu\rho}=\left(\frac{\sqrt{-\mathfrak{g}}}{\omega}\right)^{-\frac{2}{n}}\left(\delta g_{\nu\rho}-\frac{2}{n}g_{\nu\rho}\delta\ln\frac{\sqrt{-\mathfrak{g}}}{\omega}\right).
\end{equation}
Then it is easy to check that $\theta^{\mu}$ is WTDiff invariant.

In the unimodular gauge, $\sqrt{-\mathfrak{g}}=\omega$, and for variations that do not change the metric determinant, $\delta \mathfrak{g}=0$, the symplectic potential of WTG coincides with the GR result~\cite{Wald:1994},
\begin{equation}
\theta^{\mu}_{\text{GR}}=\frac{1}{16\pi}\sqrt{-\mathfrak{g}}\left(g^{\mu\nu}g^{\rho\sigma}-g^{\mu\sigma}g^{\nu\rho}\right)\nabla_{\sigma}\delta g_{\nu\rho}.
\end{equation}
This is of course expected, since the WTG action reduces to the Einstein-Hilbert action of GR in the unimodular gauge.

\paragraph{Symplectic form.}

Consider two general, independent variations of the metric, $\delta_1 g_{\mu\nu}$ and $\delta_2 g_{\mu\nu}$. The corresponding symplectic current~\eqref{omega g} equals
\begin{align}
\label{omega}
\Omega^{\mu}\left[\delta_{1},\delta_{2}\right]=&\delta_{1}\theta^{\mu}\left[\delta_{2}\right]-\delta_{2}\theta^{\mu}\left[\delta_{1}\right].
\end{align}
The symplectic form, $\Omega\left[\delta_{1},\delta_{2}\right]$ is then given by an integral of $\Omega^{\mu}\left[\delta_{1},\delta_{2}\right]$ over an appropriate initial data surface, $\mathcal{C}$.

Evaluating the symplectic form for a transverse diffeomorphism generated by a vector field $\xi^{\mu}$, $\delta_1 g_{\mu\nu}=\pounds_{\xi}g_{\mu\nu}$, and a general metric variation, $\delta_{2}g_{\mu\nu}=\delta g_{\mu\nu}$, yields the variation of the Hamiltonian corresponding to the evolution along $\xi^{\mu}$
\begin{equation}
\label{hamilton eq}
\delta H_{\xi}=\Omega\left[\pounds_{\xi},\delta\right].
\end{equation}
Obtaining an expression for $\delta H_{\xi}$ is one of our main goals. While we could apply the above described method, it requires the precise form of the Lagrangian. In the following, we discuss an alternative route applicable to Hamiltonians corresponding to generators of the symmetries of the theory~\cite{Wald:1993,Wald:1994} (in our case the WTDiff group). Although the calculations are less straightforward, they can be applied to any Lagrangian with WTDiff symmetry. This will allow us to derive the matter contributions to the perturbation of the Hamiltonian in subsection~\ref{sourced}. Furthermore, this method relates the Hamiltonian perturbations with the Noether charges corresponding to symmetry transformations. As a by-product, it also yields several relations valid even off shell.

\subsection{WTG Noether current}
\label{current}

So far, we have considered general variations of the metric. We now focus on variations that do not change the physical content of the theory, i.e., those corresponding to local symmetry transformations, and calculate the Noether currents and charges. For WTG, the local symmetry transformations are transverse diffeomorphisms, $\delta_{\xi}g_{\mu\nu}=2\nabla_{(\mu}\xi_{\nu)}$ with $\tilde{\nabla}_{\mu}\xi^{\mu}=0$, and Weyl transformations, $\delta_{\text{W}}g_{\mu\nu}=\phi g_{\mu\nu}$ for some function $\phi$.

We start with local Weyl transformations. The corresponding variation of the WTG Lagrangian vanishes, $\delta_{\text{W}}L=0$. Hence, $\alpha^{\mu}$ in the general definition of the Noether current~\eqref{j} equals zero. Furthermore, it holds $\theta^{\mu}_{\text{W}}=0$. In total, we have that the Noether current $j^{\mu}_{\text{W}}$ associated with Weyl invariance vanishes, $j^{\mu}_{\text{W}}=0$, in agreement with earlier observations~\cite{Jackiw:2014,Oda:2017}.

Let us now turn to transverse diffeomorphisms generated by a vector field $\xi^{\mu}$. The variation of the Lagrangian equals
\begin{equation}
\delta_{\xi}L=\pounds_{\xi}L=\tilde{\nabla}_{\mu}\left(\frac{1}{16\pi}\tilde{R}\xi^{\mu}\right),
\end{equation}
where we used $\tilde{\nabla}_{\mu}\xi^{\mu}=0$. Therefore, we have $\alpha^{\mu}=\tilde{R}\xi^{\mu}/16\pi$.

For the symplectic potential corresponding to a transverse diffeomorphism, we find from equation~\eqref{theta}
\begin{align}
\nonumber \theta^{\mu}\left[\pounds_{\xi}\right]=&\frac{1}{8\pi}\left(\sqrt{-\mathfrak{g}}/\omega\right)^{2/n}\left(g^{\mu\nu}g^{\rho\sigma}-g^{\mu\sigma}g^{\nu\rho}\right)g_{\lambda(\nu\vert}\tilde{\nabla}_{\sigma}\tilde{\nabla}_{\vert\rho)}\xi^{\lambda} \\
=&\frac{1}{8\pi}\tilde{g}^{\mu\rho}\tilde{R}_{\rho\nu}\xi^{\nu}+\tilde{\nabla}_{\nu}\left[\frac{1}{8\pi}\left(\sqrt{-\mathfrak{g}}/\omega\right)^{2/n}\tilde{\nabla}^{[\nu}\xi^{\mu]}\right],
\end{align}
where we repeatedly employed the commutator of Weyl covariant derivatives
\begin{equation}
\left(\tilde{\nabla}_{\sigma}\tilde{\nabla}_{\rho}-\tilde{\nabla}_{\rho}\tilde{\nabla}_{\sigma}\right)\xi^{\lambda}=-\tilde{R}^{\lambda}_{\;\:\nu\rho\sigma}\xi^{\nu}.
\end{equation}
In total, the corresponding Noether current $j^{\mu}_{\xi}=j^{\mu}\left[\pounds_{\xi}\right]$ reads
\begin{equation}
\label{j WTG}
j^{\mu}_{\xi}=\frac{1}{8\pi}\left(\tilde{g}^{\mu\rho}\tilde{R}_{\rho\nu}-\frac{1}{2}\tilde{R}\delta^{\mu}_{\nu}\right)\xi^{\nu}+\tilde{\nabla}_{\nu}\left[\frac{1}{8\pi}\left(\sqrt{-\mathfrak{g}}/\omega\right)^{2/n}\tilde{\nabla}^{[\nu}\xi^{\mu]}\right].
\end{equation}
While this form of $j^{\mu}_{\xi}$ is perfectly well defined without any reference to the equations of motion, we also often work on shell in the following. In that case, the vacuum WTG equations of motion imply 
\begin{equation}
\label{j shell}
j^{\mu}_{\xi}=-\frac{1}{8\pi}\Lambda\xi^{\mu}+\tilde{\nabla}_{\nu}\left[\frac{1}{8\pi}\left(\sqrt{-\mathfrak{g}}/\omega\right)^{2/n}\tilde{\nabla}^{[\nu}\xi^{\mu]}\right],
\end{equation}
where $\Lambda$ is an arbitrary integration constant.

For the Weyl invariant divergence of the off-shell Noether current $j^{\mu}_{\xi}$, we find
\begin{align}
\nonumber \tilde{\nabla}_{\mu}j^{\mu}_{\xi}=&\tilde{\nabla}_{\mu}\left[\frac{1}{8\pi}\left(\tilde{g}^{\mu\rho}\tilde{R}_{\rho\nu}-\frac{1}{2}\tilde{R}\delta^{\mu}_{\nu}\right)\xi^{\nu}\right]+\tilde{\nabla}_{\mu}\tilde{\nabla}_{\nu}\left[\frac{1}{8\pi}\left(\sqrt{-\mathfrak{g}}/\omega\right)^{2/n}\tilde{\nabla}^{[\nu}\xi^{\mu]}\right] \\
=&\frac{1}{8\pi}\left(\sqrt{-\mathfrak{g}}/\omega\right)^{2/n}\left(\tilde{R}_{\mu\nu}-\frac{1}{2}\tilde{R}\tilde{g}_{\mu\nu}\right)\tilde{\nabla}^{\nu}\xi^{\mu},
\end{align}
where we modified the first term using the contracted Bianchi identities~\eqref{Bianchi} and the second term drops out due to antisymmetrisation. On shell, it holds
\begin{equation}
\tilde{\nabla}_{\mu}j^{\mu}_{\xi}=-\frac{1}{8\pi}\Lambda\tilde{\nabla}_{\mu}\xi^{\mu}=0,
\end{equation}
and the divergence of the Noether current vanishes by the transversality condition.

The second term in the Noether current~\eqref{j WTG} corresponds to the divergence of the Noether charge antisymmetric tensor, $Q^{\nu\mu}_{\xi}=Q^{[\nu\mu]}_{\xi}$, which equals
\begin{equation}
\label{charge}
Q^{\nu\mu}_{\xi}=\frac{1}{8\pi}\left(\frac{\sqrt{-\mathfrak{g}}}{\omega}\right)^{\frac{2}{n}}\tilde{\nabla}^{[\nu}\xi^{\mu]}.
\end{equation}
One can easily check that $Q^{\nu\mu}_{\xi}$ is WTDiff invariant.

As expected, in the unimodular gauge the WTG Noether charge coincides with the corresponding GR expression
\begin{equation}
Q^{\nu\mu}_{\text{GR},\xi}=\frac{1}{8\pi}\sqrt{-\mathfrak{g}}\nabla^{[\nu}\xi^{\mu]}.
\end{equation}
However, the WTG Noether current in the unimodular gauge becomes
\begin{equation}
j^{\mu}_{\xi}=\frac{1}{8\pi}\sqrt{-\mathfrak{g}}\left(R_{\nu}^{\;\:\mu}-\frac{1}{2}R\delta_{\nu}^{\mu}\right)\xi^{\rho}+\frac{1}{8\pi}\sqrt{-\mathfrak{g}}\nabla_{\nu}\nabla^{[\nu}\xi^{\mu]},
\end{equation}
and, compared to the Noether current associated with transverse diffeomorphisms in GR
\begin{equation}
j^{\mu}_{\text{GR},\xi}=\frac{1}{8\pi}\sqrt{-\mathfrak{g}}\left(R_{\nu}^{\;\:\mu}-\frac{1}{2}R\delta_{\nu}^{\mu}+\Lambda\delta_{\nu}^{\mu}\right)\xi^{\rho}+\frac{1}{8\pi}\sqrt{-\mathfrak{g}}\nabla_{\nu}\nabla^{[\nu}\xi^{\mu]},
\end{equation}
it misses a term proportional to $\Lambda$. This difference is just a divergenceless contribution. To explain it, note that the Einstein-Hilbert Lagrangian for GR, $L_{\text{EH}}=\left(R-2\Lambda\right)/16\pi$, contains $\Lambda$ as a constant fixed parameter, whereas the WTG Lagrangian, $L=\tilde{R}/16\pi$, does not. This is the origin of the discrepancy, since the Noether current contains a term $-\alpha^{\mu}=-L\xi^{\mu}$ in both theories.

Before going further, we should mention (and dismiss) an apparent way to match the WTG and GR Noether currents. As we showed in equation~\ref{action}, the WTG Lagrangian may in principle include a term of the form $-\lambda/8\pi$, with $\lambda$ being an arbitrary constant. As $\lambda$ does not affect the gravitational dynamics in any way, we set $\lambda=0$. However, if we choose nonzero $\lambda$ in the Lagrangian, it shifts the WTG Noether current by $\lambda\xi^{\mu}/8\pi$. Since we have $\tilde{\nabla}_{\mu}\xi^{\mu}=0$, the Weyl divergence of the shifted Noether current still vanishes on shell. One might be tempted to use this freedom to set $\lambda=\Lambda$ and recover the GR form of the Noether current in the unimodular gauge. However, as $\Lambda$ in WTG appears as an integration constant in the process of solving the equations of motion, it is well defined only on shell and in general different for each solution. Hence, the single constant $\lambda$ which we are free to choose at the level of the Lagrangian cannot match all the different values of $\Lambda$ for various solutions of the equations of motion. In the following, we continue to assume $\lambda=0$, as it it is completely irrelevant for our considerations (we briefly return to its interpretation in subsection~\ref{lambda}).

Lastly, we note that the above defined Noether current and charge, as well as the symplectic potential and current contain some ambiguities. Since these are the same as for the Noether charge formalism of Diff invariant gravity~\cite{Wald:1994}, we do not discuss them in detail. In any case, they do not affect the physical situations we investigate in this paper. Hence, in the following, we simply treat all the expressions derived up to this point as effectively unambiguous.

\subsection{WTG Hamiltonian for transverse diffeomorphisms}
\label{hamiltonian}

The Noether current corresponding to a transverse diffeomorphism can be used to obtain the Hamiltonian for evolution along its generator, $\xi^{\mu}$. To see this, consider any variation of the metric, $\delta g_{\mu\nu}$. We have two ways of calculating the corresponding variation of $j_{\xi}^{\mu}$. On one side, the general definition of $j_{\xi}^{\mu}$ with $\alpha^{\mu}=L\xi^{\mu}$ implies
\begin{align}
\nonumber \delta j^{\mu}_{\xi}=\delta\theta^{\mu}\left[\pounds_{\xi}\right]-\xi^{\mu}\delta L.
\end{align}
Using equation~\eqref{dL} to express the variation of the WTG Lagrangian yields
\begin{equation}
\label{dj0}
\delta j^{\mu}_{\xi}=\delta\theta^{\mu}\left[\pounds_{\xi}\right]-\xi^{\mu}\tilde{\nabla}_{\nu}\theta^{\nu}\left[\delta\right]-\frac{1}{16\pi}\xi^{\mu}\tilde{g}^{\alpha\rho}\tilde{g}^{\beta\sigma}\left(\tilde{R}_{\rho\sigma}-\frac{1}{n}\tilde{R}\tilde{g}_{\rho\sigma}\right)\delta g_{\alpha\beta}.
\end{equation}
To obtain the variation of the Hamiltonian given in general by equation~\eqref{dH0}, we first need to identify the symplectic current~\eqref{omega} corresponding to a transverse diffeomorphism generated by $\xi^{\mu}$ and an arbitrary variation of the metric,
\begin{equation}
\Omega^{\mu}\left[\pounds_{\xi},\delta\right]=\delta\theta^{\mu}\left[\pounds_{\xi}\right]-\pounds_{\xi}\theta^{\mu}\left[\delta\right].
\end{equation}
The first term is already present in equation~\eqref{dj0}. If we add and subtract the second term, $\pounds_{\xi}\theta^{\mu}\left[\delta\right]$, to equation~\eqref{dj0} and use that it holds
\begin{equation}
\label{lie theta}
\pounds_{\xi}\theta^{\mu}=\xi^{\nu}\tilde{\nabla}_{\nu}\theta^{\mu}-\theta^{\nu}\tilde{\nabla}_{\nu}\xi^{\mu},
\end{equation}
we obtain
\begin{equation}
\label{dj4}
\delta j^{\mu}_{\xi}=\Omega^{\mu}\left[\pounds_{\xi},\delta\right]+2\tilde{\nabla}_{\nu}\left(\xi^{[\nu}\theta^{\mu]}\left[\delta\right]\right)-\frac{1}{16\pi}\left(\sqrt{-\mathfrak{g}}/\omega\right)^{2/n}\xi^{\mu}g^{\alpha\rho}g^{\beta\sigma}\left(\tilde{R}_{\rho\sigma}-\frac{1}{n}\tilde{R}\tilde{g}_{\rho\sigma}\right)\delta g_{\alpha\beta}.
\end{equation}

On the other side, we expressed the Noether current in terms of the equations of motion and the Noether charge~\eqref{charge}. A variation of this expression reads
\begin{equation}
\label{dj5}
\delta j^{\mu}_{\xi}=\frac{1}{8\pi}\xi^{\nu}\delta\left(\tilde{g}^{\mu\rho}\tilde{R}_{\rho\nu}-\frac{1}{2}\tilde{R}\delta^{\mu}_{\nu}\right)+\tilde{\nabla}_{\nu}\delta Q^{\nu\mu}_{\xi}.
\end{equation}
Equating both expressions~\eqref{dj4} and~\eqref{dj5} for $\delta j^{\mu}_{\xi}$ yields the desired expression for the symplectic current
\begin{align}
\label{dj}
\nonumber \Omega^{\mu}\left[\pounds_{\xi},\delta\right]=&\tilde{\nabla}_{\nu}\left(Q^{\nu\mu}_{\xi}-2\xi^{[\nu}\theta^{\mu]}\left[\delta \right]\right)+\frac{1}{8\pi}\xi^{\nu}\delta\left(\tilde{g}^{\mu\rho}\tilde{R}_{\rho\nu}-\frac{1}{2}\tilde{R}\delta^{\mu}_{\nu}\right) \\
&+\frac{1}{16\pi}\left(\sqrt{-\mathfrak{g}}/\omega\right)^{2/n}\xi^{\mu}g^{\alpha\rho}g^{\beta\sigma}\left(\tilde{R}_{\rho\sigma}-\frac{1}{n}\tilde{R}\tilde{g}_{\rho\sigma}\right)\delta g_{\alpha\beta}.
\end{align}
Note that equation~\eqref{dj} has been derived without requiring that the equations of motion hold.

Now assume that we compare two solutions of the vacuum WTG equations of motion related by a small perturbation. Then, the perturbation obeys the equations of motion, 
\begin{equation}
\delta\left(\tilde{g}^{\mu\rho}\tilde{R}_{\rho\nu}-\frac{1}{2}\tilde{R}\delta^{\mu}_{\nu}\right)=-\delta\Lambda\delta^{\mu}_{\nu},
\end{equation}
where $\delta\Lambda$ is an arbitrary integration constant. Clearly, the cosmological constant in the perturbed spacetime equals $\Lambda+\delta\Lambda$. In this way, our formalism naturally allows for $\Lambda$ to differ between various solutions of the equations of motion. The symplectic current then considerably simplifies on shell
\begin{equation}
\label{omega 2}
\Omega^{\mu}\left[\pounds_{\xi},\delta\right]=\tilde{\nabla}_{\nu}\left(\delta Q^{\nu\mu}_{\xi}-2\xi^{[\nu}\theta^{\mu]}\left[\delta\right]\right)-\frac{1}{8\pi}\xi^{\mu}\delta\Lambda.
\end{equation}

Suppose that the unperturbed spacetime possesses a Cauchy surface, $\mathcal{C}$. We introduce a Weyl invariant volume element on $\mathcal{C}$, $\text{d}\mathcal{C}_{\mu}=\left(\sqrt{-\mathfrak{g}}/\omega\right)^{-1/n}n_{\mu}\omega\text{d}^{n-1}x$, where $n_{\mu}$ denotes a unit normal to $\mathcal{C}$ and $\text{d}^{n-1}x$ is the coordinate volume element on $\mathcal{C}$. We stress that $\text{d}\mathcal{C}_{\mu}$ only reduces to the physical volume element (measured with respect to the dynamical metric) in the unimodular gauge, $\sqrt{-\mathfrak{g}}=\omega$. Integrating equation~\eqref{omega 2} over $\mathcal{C}$ with this volume element and rewriting the integral of the first term as an integral over the boundary $\partial\mathcal{C}$ of $\mathcal{C}$ yields a Weyl invariant expression for the symplectic form
\begin{equation}
\label{symplectic}
\Omega\left[\pounds_{\xi},\delta\right]=\int_{\mathcal{C}}\Omega^{\mu}\left[\pounds_{\xi},\delta\right]\text{d}\mathcal{C}_{\mu}=\int_{\partial\mathcal{C}}\left(\delta Q^{\nu\mu}_{\xi}-2\xi^{\nu}\theta^{\mu}\left[\delta\right]\right)\text{d}\mathcal{C}_{\mu\nu}-\int_{\mathcal{C}}\frac{1}{8\pi}\delta\Lambda\xi^{\mu}\text{d}\mathcal{C}_{\mu},
\end{equation}
where $\text{d}\mathcal{C}_{\mu\nu}=\left(\sqrt{-\mathfrak{g}}/\omega\right)^{-2/n}n_{[\mu}m_{\nu]}\omega\text{d}^{n-2}x$ denotes the Weyl invariant area element on the boundary $\partial\mathcal{C}$ (with $m_{\mu}$ being the unit normal to $\partial\mathcal{C}$ with respect to its embedding in $\mathcal{C}$ and $\text{d}^{n-2}x$ being the coordinate area element). Provided that the Hamiltonian $H_{\xi}$ corresponding to the evolution along $\xi^{\mu}$ exists, the Hamilton equations of motion imply
\begin{equation}
\label{dH1}
\delta H_{\xi}=\Omega\left[\pounds_{\xi},\delta\right]=\int_{\partial\mathcal{C}}\left(\delta Q^{\nu\mu}_{\xi}-2\xi^{\nu}\theta^{\mu}\left[\delta\right]\right)\text{d}\mathcal{C}_{\mu\nu}-\int_{\mathcal{C}}\frac{1}{8\pi}\delta\Lambda\xi^{\mu}\text{d}\mathcal{C}_{\mu}.
\end{equation}
In contrast with the situation in fully diffeomorphism invariant theories, the Hamiltonian no longer consists only of surface terms~\cite{Wald:1993}. Instead, it includes a volume integral proportional to the perturbation of the cosmological constant. This is a consequence of the fact that $\Lambda$ is an integration constant rather than a fixed parameter in the WTG Lagrangian. Since the contribution of the volume integral is clearly infinite, cases with nonzero $\Lambda$ or $\delta\Lambda$ require a separate treatment, which we discuss in subsection~\ref{lambda}.

For $\Lambda=0$, i.e., when the Hamiltonian $H_{\xi}$ is finite, the necessary and sufficient condition for its existence is the same as for diffeomorphism invariant theories~\cite{Wald:2000}
\begin{equation}
\label{condition}
\int_{\partial\mathcal{C}}\Omega^{\mu}\left[\pounds_{\xi},\delta\right]\xi^{\nu}\text{d}\mathcal{C}_{\mu\nu}=0.
\end{equation}
To see this, consider some solution of the vacuum WTG equations with $\Lambda=0$. On this solution, choose a vector field $\xi^{\mu}$, so that the Hamiltonian $H_{\xi}$ exists. We consider two independent variations of the metric, $\delta_1g_{\mu\nu}$ and $\delta_2g_{\mu\nu}$, which satisfy the equations of motion with $\delta_1\Lambda=\delta_2\Lambda=0$. Then, it holds $\left(\delta_1\delta_2-\delta_2\delta_1\right)H_{\xi}=0$. Using equation~\eqref{dH1} for the variation of the Hamiltonian, the commutation of $\delta_1$ and $\delta_2$ on the antisymmetric tensor $Q^{\nu\mu}_{\xi}$, and the definition of $\Omega^{\mu}\left[\pounds_{\xi},\delta\right]$, we get the condition~\eqref{condition}.

\section{The first law and Wald entropy in vacuum}
\label{physics}

In the previous section, we have obtained an expression for the perturbation of a Hamiltonian corresponding to evolution along a transverse diffeomorphism generator, $\xi^{\mu}$. In the following, we use it to study several special cases of interest. Given the classical equivalence of WTG and GR, we naturally expect an agreement of our results with the ones known for GR. Even so, the machinery we developed presents some advantages. First, it allows us to treat situations in which the energy-momentum tensor is not locally conserved, i.e., $\mathcal{J}\ne0$. Second, as $\Lambda$ appears as an integration constant in WTG, it can vary between solutions. Therefore, our formalism allows for a very natural treatment of contributions from a varying cosmological constant to the first law, a possibility studied in the context of black hole chemistry~\cite{Kubiznak:2015}.

We first discuss an asymptotically flat, vacuum black hole spacetime and define the Wald entropy for WTG. Then we turn to two situations with a nonzero cosmological constant, an asymptotically anti-de Sitter black hole and de Sitter spacetime.

\subsection{WTG first law of black hole mechanics}
\label{black holes}

In this subsection we concentrate on spacetimes with $\Lambda=0$. We will focus on nonzero $\Lambda$ in subsection~\ref{lambda}. We start by discussing a general case of an asymptotically flat spacetime which possesses a Cauchy surface, $\mathcal{C}$, and a Killing vector, $\xi^{\mu}$. We adopt a WTDiff invariant definition of a Killing vector, i.e., $\tilde{g}_{\rho(\nu}\tilde{\nabla}_{\mu)}\xi^{\rho}=0$. This implies $\pounds_{\xi}\tilde{g}_{\mu\nu}=0$. For the dynamical metric, we have $\pounds_{\xi}g_{\mu\nu}=g_{\mu\nu}\xi^{\rho}\partial_{\rho}\ln\left(\sqrt{-\mathfrak{g}}/\omega\right)$. Hence, the change of $g_{\mu\nu}$ corresponds to a pure Weyl transformation. This definition of a Killing vector agrees with the standard one, $\nabla_{(\mu}\xi_{\nu)}=0$, in the unimodular gauge, in which equations of motion for WTG and GR coincide. Beyond this gauge, the WTDiff invariant definition ensures that the Killing vectors of any two spacetimes related by a Weyl transformations will be the same (this is desirable, since such spacetimes are physically identical in WTG). We stress that even though the Killing vectors do not affect $\tilde{g}_{\mu\nu}$, this statements depends on the specific form of the metric. Hence, the Noether current and charge derived for general metrics are generically nonvanishing when evaluated for a Killing vector.

In this spacetime, we consider an arbitrary perturbation of the metric which obeys the equations of motion. We further require that the metric perturbation does not spoil the asymptotic flatness, in particular, we impose $\delta\Lambda=0$.
Since the perturbation of the Hamiltonian corresponding to the evolution along the Killing vector $\xi^{\mu}$ vanishes (as $\pounds_{\xi}\tilde{g}_{\mu\nu}=0$ for any Killing vector, the symplectic current $\Omega^{\mu}\left[\pounds_{\xi},\delta\right]$~\eqref{omega} identically vanishes and so does $\delta H_{\xi}$ according to equation~\eqref{symplectic}), equation~\eqref{dH1} yields
\begin{equation}
\label{dH2}
\delta H_{\xi}=\int_{\partial\mathcal{C}}\left(\delta Q^{\nu\mu}_{\xi}-2\xi^{\nu}\theta^{\mu}\left[\delta\right]\right)\text{d}\mathcal{C}_{\mu\nu}=0.
\end{equation}
The boundary $\partial\mathcal{C}$ of the Cauchy surface $\mathcal{C}$ has in general several components. One is an intersection $\partial\mathcal{C}_{\infty}$ of the Cauchy surface $\mathcal{C}$ with the spatial infinity\footnote{Here and in the following, we consider a causal structure defined with respect to the background metric, $\tilde{g}_{\mu\nu}$. This is a consequence of the WTDiff invariance of physics in WTG.}. There can be further internal components of the boundary, e.g. intersections of $\mathcal{C}$ with black hole horizons. We collectively denote these by $\partial\mathcal{C}_{\text{I}}$.  Then, equation~\eqref{dH2} can be written as
\begin{equation}
\label{dH killing}
\int_{\partial\mathcal{C}_{\infty}}\left(\delta Q^{\nu\mu}_{\xi}-2\xi^{\nu}\theta^{\mu}\left[\delta\right]\right)\text{d}\mathcal{C}_{\mu\nu}-\int_{\partial\mathcal{C}_{\text{I}}}\left(\delta Q^{\nu\mu}_{\xi}-2\xi^{\nu}\theta^{\mu}\left[\delta\right]\right)\text{d}\mathcal{C}_{\mu\nu}=0.
\end{equation}
The interpretation of the second term depends on the nature of $\partial\mathcal{C}_{\text{I}}$. The first term corresponds to the perturbations of quantities measured in the asymptotic infinity. For example, suppose that a spacetime possesses a timelike Killing vector, $t^{\mu}$, normalised so that $\tilde{g}_{\mu\nu}t^{\mu}t^{\nu}=-1$ in the asymptotic infinity. Then, the contribution to $\delta H_{t}$ from $\partial\mathcal{C}_{\infty}$ can be identified with the perturbation of the total canonical energy of the spacetime,
\begin{equation}
\label{energy}
\delta E=\int_{\partial\mathcal{C}_{\infty}}\left(\delta Q^{\mu\nu}_{t}-2t^{\nu}\theta^{\mu}\left[\delta\right]\right)\text{d}\mathcal{C}_{\mu\nu}.
\end{equation}
Similarly, the presence of a rotational Killing vector $\varphi^{\mu}$ allows us to define the perturbation of the total canonical angular momentum
\begin{equation}
\label{angular}
\delta J=-\int_{\partial\mathcal{C}_{\infty}}\delta Q^{\mu\nu}_{\varphi}\text{d}\mathcal{C}_{\mu\nu},
\end{equation}
where the overall minus sign ensures that $J$ is positive. Note that, since $\varphi^{\mu}$ is orthogonal to $\text{d}\mathcal{C}_{\mu\nu}$, there is no contribution proportional to $\varphi^{\nu}\theta^{\mu}$. Both expressions are explicitly Weyl invariant. In the unimodular gauge and for perturbations preserving the metric determinant, $\delta\mathfrak{g}=0$, they reduce to the perturbations of the familiar ADM energy and angular momentum for GR.

We now specialise to the case of a stationary spacetime with a single black hole. Any such spacetime is endowed with a time translational Killing vector $t^{\mu}$ and a set of $n-3$ Killing vector fields $\varphi^{\mu}_{(i)}$ corresponding to rotations. These can be combined into a Killing vector field $\xi^{\mu}=t^{\mu}+\sum_{i=1}^{n-3}\Omega_{\mathcal{H}}^{(i)}\varphi^{\mu}_{(i)}$, where $\Omega_{\mathcal{H}}^{(i)}$ are constant angular velocities of the black hole horizon in directions $\varphi^{\mu}_{(i)}$. The black hole's horizon is then a bifurcate Killing horizon with respect to $\xi^{\mu}$. A spacelike Cauchy surface $\mathcal{C}$ has an inner boundary formed by the intersection of $\mathcal{C}$ with the black hole's event horizon, $\mathcal{H}$, which we denote by $\partial\mathcal{C}_{\mathcal{H}}$. Applying equation~\eqref{dH killing} to this case yields
\begin{equation}
\delta E-\sum_{i=1}^{n-3}\Omega^{(i)}_{\mathcal{H}}\delta J_{(i)}-\frac{1}{8\pi}\int_{\partial\mathcal{C}_{\mathcal{H}}}\left[\delta\left(\left(\sqrt{-\mathfrak{g}}/\omega\right)^{2/n}\tilde{\nabla}^{\nu}\xi^{\mu}\right)-2\xi^{\nu}\theta^{\mu}\right]\text{d}\mathcal{C}_{\mu\nu}=0,
\end{equation}
where we identified the perturbations of energy~\eqref{energy} and angular momenta~\eqref{angular}. The Killing vector $\xi^{\mu}$ is normal to $\mathcal{H}$. Hence, the second term in the integral over $\partial\mathcal{C}_{\mathcal{H}}$ vanishes. If we define the Weyl invariant surface gravity of the horizon,
\begin{equation}
\label{kappa}
\kappa=\sqrt{g_{\mu\nu}g^{\rho\sigma}\tilde{\nabla}_{\rho}\xi^{\mu}\tilde{\nabla}_{\sigma}\xi^{\nu}}\Big\vert_{\mathcal{H}},
\end{equation}
it is easy to show that $\tilde{\nabla}^{\nu}\xi^{\mu}=\kappa\epsilon^{\nu\mu}$, where $\epsilon^{\nu\mu}$ is the bi-normal to the horizon. Finally, we obtain
\begin{equation}
\label{first law}
\delta E-\sum_{i=1}^{n-3}\Omega^{(i)}_{\mathcal{H}}\delta J_{(i)}-\frac{1}{8\pi}\kappa\int_{\partial\mathcal{C}_{\mathcal{H}}}\delta\left[\left(\sqrt{-\mathfrak{g}}/\omega\right)^{2/n}\epsilon^{\nu\mu}\right]\text{d}\mathcal{C}_{\mu\nu}=0.
\end{equation}
In the unimodular gauge and for perturbations which do not change the metric determinant, this reduces to the standard first law of black hole mechanics known for GR~\cite{Wald:1994}
\begin{equation}
\delta E-\sum_{i=1}^{n-3}\Omega^{(i)}_{\mathcal{H}}\delta J_{(i)}-\frac{1}{8\pi}\kappa\delta\mathcal{A}=0,
\end{equation}
where $\mathcal{A}$ denotes the area of the horizon's cross-section $\partial\mathcal{C}_{\mathcal{H}}$. It is then proved that the formulations of the first law in WTG and GR are physically equivalent.

\subsection{Wald entropy}

The previous subsection concerns purely classical gravity. That suffices to derive the first law of black hole mechanics. However, to identify black hole entropy and promote the first law into a genuine thermodynamic statement, we require insights from the quantum field theory in a curved background. It is well known that due to quantum effects black holes emit black body radiation (Hawking radiation) corresponding to a finite temperature~\cite{Hawking:1975}. Hawking radiation results from fluctuations of the matter fields, which are Weyl invariant (and there are no quantum anomalies associated with local Weyl transformations~\cite{Carballo:2015,Alvarez:2013b}). Hence, we expect the Hawking temperature to be Weyl invariant as well. Since the Hawking radiation is a kinematic effect independent of the gravitational dynamics~\cite{Visser:2003}, the standard expression for the Hawking temperature, $T_{\text{H}}=\kappa/2\pi$, must hold. We just need to specify $\kappa$ to be the Weyl invariant surface gravity~\eqref{kappa} to ensure the overall Weyl invariance of $T_{\text{H}}$.

Hawking radiation allows us to identify a term of the form $T_{\text{H}}\delta S$ in the first law and define Wald entropy of the horizon, $S$. For the bifurcate Killing horizon of a stationary black hole it then holds
\begin{equation}
S=\frac{1}{4}\int_{\partial\mathcal{C}_{\mathcal{H}}}\left(\sqrt{-\mathfrak{g}}/\omega\right)^{2/n}\epsilon^{\nu\mu}\text{d}\mathcal{C}_{\mu\nu}=\frac{1}{4}\int_{\partial\mathcal{C}_{\mathcal{H}}}\left(\sqrt{-\mathfrak{g}}/\omega\right)^{\left(2-n\right)/n}\sqrt{\mathfrak{h}}\text{d}^{n-2}x,
\end{equation}
with $\mathfrak{h}$ being the determinant of the $\left(n-2\right)$-dimensional reduced metric on $\partial\mathcal{C}_{\mathcal{H}}$ and $\text{d}^{n-2}x$ is the corresponding coordinate area element. As expected, defining the Weyl invariant Hawking temperature also ensures the Weyl invariance of Wald entropy. In the unimodular gauge, the WTG Wald entropy reduces to the well-known Bekenstein entropy of GR, \mbox{$S_{\text{B}}=\mathcal{A}/4$}.

Defining $T_{\text{H}}$ and $S$ in this way yields the following first law of black hole thermodynamics for stationary, asymptotically flat black holes in vacuum
\begin{equation}
0=\delta E-\sum_{i=1}^{n-3}\Omega^{(i)}_{\mathcal{H}}\delta J_{(i)}-T_{\text{H}}\delta S,
\end{equation}
which has the form familiar from GR. However, the total energy, angular momenta, Hawking temperature, and Wald entropy are all Weyl invariant and only agree with the corresponding GR expressions in the unimodular gauge.

Finally, let us once again stress that the Hawking temperature, rather than emerging naturally in the Noether charge formalism, requires an additional insight from the quantum field theory in a curved background. We have just introduced it in this subsection to find Wald entropy of WTG for the sake of comparison with GR. In the rest of the paper, we return to considering only classical physics.

\subsection{Spacetimes with non-zero cosmological constant}
\label{lambda}

Since the main difference between GR and WTG lies in the nature of the cosmological constant, it is of interest to apply the WTDiff invariant Noether charge formalism to spacetimes with $\Lambda\ne0$. We do so for two physically interesting yet tractable examples, a Schwarzschild-anti-de Sitter black hole and de Sitter spacetime in four spacetime dimensions. Applying our formalism to more general spacetimes with a non-zero cosmological constant, e.g. stationary, asymptotically (anti-)de Sitter black holes, would be more or less straightforward. However, we choose the simplest cases to more clearly illustrate the peculiar features of the cosmological constant in WTG.

\paragraph{Schwarzschild-anti-de Sitter spacetime.}

The simplest way to find a solution of the WTG equations of motion corresponding to a known solution of the Einstein equations is by expressing the metric in the unimodular gauge (using a suitable diffeomorphism). Since the equations of motion of WTG and GR coincide in this gauge, the metric is then also a solution in WTG. The four-dimensional ($n=4$) Schwarzschild-anti-de Sitter spacetime can be described by the following unimodular ($\sqrt{-\mathfrak{g}}=\omega$) metric
\begin{equation}
\label{S-AdS}
\text{d}s^2=\sqrt{\omega}\left[-\left(1-\frac{2M}{r}-\frac{\Lambda}{3}r^2\right)\text{d}t^2+\frac{\text{d}R^2}{r^4\left(1-\frac{2M}{r}-\frac{\Lambda}{3}r^2\right)}+r^2\left(\frac{\text{d}x^2}{\left(1-x^2\right)}+\left(1-x^2\right)\text{d}\phi^2\right)\right],
\end{equation}
where  $\Lambda<0$ is the cosmological constant, and, with respect to the usual spherical coordinates $r$, $\theta$, $\phi$, we defined $R=r^3/3$ and $x=-\cos\theta$ (thus, $x\in\left[-1,1\right)$). We choose this modification of the standard Schwarzschild coordinates to obtain a unimodular metric (while there exist other unimodular forms of the Schwarzschild-anti-de Sitter metric, this one is computationally convenient and also easily generalises to other black hole spacetimes). The spacetime possesses a time translational Killing vector field, $t^{\mu}=(1,0,0,0)$, and the black hole's event horizon is a Killing horizon with respect to it. Now consider equation~\eqref{j shell} for the on-shell Noether current and integrate it over a Cauchy surface orthogonal to $t^{\mu}$
\begin{equation}
\label{smarr0}
\int_{\mathcal{C}}j^{\mu}_{t}\text{d}\mathcal{C}_{\mu}=-\frac{1}{8\pi}\Lambda\int_{\mathcal{C}}t^{\mu}\text{d}\mathcal{C}_{\mu}+\int_{\partial\mathcal{C}_{\infty}}Q^{\nu\mu}_{t}\text{d}\mathcal{C}_{\mu\nu}-\int_{\partial\mathcal{C}_{\mathcal{H}}}Q^{\nu\mu}_{t}\text{d}\mathcal{C}_{\mu\nu}
\end{equation}
where we used the Gauss theorem and the division of the boundary of $\mathcal{C}$ into its intersections with the asymptotic infinity, $\partial\mathcal{C}_{\infty}$, and the black hole horizon, $\partial\mathcal{C}_{\mathcal{H}}$.

If $\Lambda=0$, then $\tilde{R}=0$ and the Lagrangian vanishes. Moreover, as $t^{\mu}$ is a Killing vector and $\pounds_{t}\tilde{g}_{\mu\nu}=0$, the corresponding symplectic potential vanishes (see equation~\eqref{theta}), and we have $j^{\mu}_{t}=\theta^{\mu}_{t}-Lt^{\mu}=0$. The integral of $Q^{\nu\mu}_{t}$ over the asymptotic infinity is in this case proportional to the total mass, $M$, and the integral over the horizon is proportional to the area. Equation~\eqref{smarr0} then becomes the Smarr formula relating the horizon area, $\mathcal{A}$, and mass $M$ of a Schwarzschild black hole
\begin{equation}
0=\frac{1}{2}M-\frac{\kappa}{8\pi}\mathcal{A},\qquad\qquad\left(\Lambda=0\right)
\end{equation}
where $\kappa$ is the surface gravity.

For $\Lambda<0$ both sides of equation~\eqref{smarr0} are infinite. However, these infinities are of the same nature as in the pure anti-de Sitter spacetime. We can then choose anti-de Sitter spacetime as our reference background and demand that the Noether current and charge vanish there. In other words, we define the physical Noether charge and current as the difference of their Schwarzschild-anti-de Sitter and pure anti de-Sitter values\footnote{This trick is similar to obtaining a finite entropy in GR by subtracting the action of a reference static background~\cite{Hawking:1996}. However, subtraction on the level of the action is problematic in WTG, since the cosmological constant is defined only on shell.},
\begin{align}
Q^{\nu\mu}_{t,\text{phys}}=&Q^{\nu\mu}_{t}-Q^{\nu\mu}_{t,\text{AdS}}, \\
j^{\mu}_{t,\text{phys}}=&j^{\mu}_{t}-j^{\mu}_{t,\text{AdS}}=\tilde{\nabla}_{\nu}Q^{\nu\mu}_{t,\text{phys}}.
\end{align}
The Smarr formula stated in terms of these physical quantities yields a finite result
\begin{align}
\nonumber \int_{\mathcal{C}}j^{\mu}_{t,\text{phys}}\text{d}\mathcal{C}_{\mu}=&\int_{\partial\mathcal{C}_{\infty}}Q^{\nu\mu}_{t,\text{phys}}\text{d}\mathcal{C}_{\mu\nu}-\int_{\partial\mathcal{C}_{\mathcal{H}}}Q^{\nu\mu}_{t,\text{phys}}\text{d}\mathcal{C}_{\mu\nu}, \\
0=&\frac{1}{2}M-\frac{1}{8\pi}\kappa\mathcal{A}_{\mathcal{H}\cap\mathcal{C}}+\frac{1}{3}\Lambda r_{\mathcal{H}}^3,
\end{align}
where $r_{\mathcal{H}}$ denotes the horizon radius. This result agrees with the Smarr formula for a Schwarzschild-anti-de Sitter black hole valid in GR~\cite{Kastor:2009}.

To obtain the first law for a Schwarzschild-anti-de Sitter black hole, we can consider a change of the physical Hamiltonian corresponding to evolution along $t^{\mu}$ between two Schwarzschild-anti-de Sitter black holes related by a small perturbation. Since the cosmological constant is allowed to vary between solutions in WTG, we must first subtract the anti-de Sitter background corresponding to the unperturbed cosmological constant $\Lambda$ from the original Schwarzschild-anti-de Sitter spacetime and the anti-de Sitter background corresponding to $\Lambda+\delta\Lambda$ from the perturbed Schwarzschild-anti-de Sitter spacetime. Only then can we calculate their difference. The resulting formula for the perturbation of the physical Hamiltonian reads
\begin{equation}
\label{dH phys}
\delta H_{t,\text{phys}}=\delta H_{t}-\delta H_{t,\text{AdS}}=\int_{\partial\mathcal{C}}\left(\delta Q^{\nu\mu}_{t,\text{phys}}-2t^{\nu}\theta^{\mu}_{\text{phys}}\left[\delta\right]\right)\text{d}\mathcal{C}_{\mu\nu},
\end{equation}
where
\begin{equation}
\theta^{\mu}_{\text{phys}}=\theta^{\mu}-\theta^{\mu}_{\text{AdS}}.
\end{equation}
Plugging the Schwarzschild-anti-de Sitter metric~\eqref{S-AdS} into the general expressions for the Noether charge and symplectic potential, subtracting the anti-de Sitter background, and evaluating the integral in equation~\eqref{dH phys} yields the WTG first law of black hole mechanics for a Schwarzschild-anti-de Sitter spacetime
\begin{equation}
\label{enthalpy}
0=\delta M-\frac{1}{8\pi}\kappa\int_{\partial\mathcal{C}_{\mathcal{H}}}\delta\left[\left(\sqrt{-\mathfrak{g}}/\omega\right)^{2/n}\epsilon^{\nu\mu}\right]\text{d}\mathcal{C}_{\mu\nu}+\frac{4\pi}{3}r_{\mathcal{H}}^3\frac{\delta\Lambda}{8\pi}.
\end{equation}
If the metric perturbation does not change the determinant, $\delta\mathfrak{g}=0$, we recover the form of the first law for a Schwarzschild-anti-de Sitter spacetime in GR~\cite{Kastor:2009} (since our background metric is chosen to be unimodular)
\begin{equation}
0=\delta M-\frac{\kappa}{8\pi}\mathcal{A}+\frac{4\pi}{3}r_{\mathcal{H}}^3{\delta\Lambda}{8\pi}.
\end{equation}
However, a perturbation of the cosmological constant $\delta\Lambda$ appears naturally in WTG, whereas it needs to be added somewhat ad hoc for GR~\cite{Kastor:2009,Jacobson:2019}, since $\Lambda$ is understood as a constant fixed parameter in the Lagrangian.

The consequences of the varying negative cosmological constant has been studied in the context of the so-called black hole chemistry~\cite{Kubiznak:2015,Kastor:2009}. It has been shown that a varying $\Lambda$ in an asymptotically anti-de Sitter black hole spacetime behaves like an effective pressure. This can be seen from the first law for a Schwarzschild-anti de Sitter black hole~\eqref{enthalpy} which contains a term of the form $\left(4\pi r_{\mathcal{H}}^3/3\right)\delta\Lambda/8\pi$. The term $\mathcal{V}_{\text{T}}=4\pi r_{\mathcal{H}}^3/3$ may be understood as a thermodynamic volume of the black hole~\cite{Kubiznak:2015}. Defining pressure associated with $\Lambda$ as $p_{\Lambda}=-\Lambda/8\pi$ gives us a term of the form $-\mathcal{V}_{\text{T}}\delta p_{\Lambda}$. If we further invoke the Hawking effect and view the second term in equation~\eqref{enthalpy} as $T_{\text{H}}\delta S$, it becomes a genuine first law of thermodynamics. The total mass of the black hole, $M$, then obeys \mbox{$\delta M=T_{\text{H}}\delta S-\mathcal{V}_{\text{T}}\delta p_{\Lambda}$}. Therefore, it corresponds to the enthalpy of the system~\cite{Kastor:2009,Kubiznak:2015}. This picture allows to describe the behaviour of asymptotically anti-de Sitter black holes in a standard thermodynamic language. Among other insights, it led to to the notion of black hole phase transitions~\cite{Kubiznak:2015}.

\paragraph{De Sitter spacetime.}

The situation in asymptotically de Sitter spacetimes is somewhat more complicated~\cite{Jager:2008}. Therefore, in the present work, we limit ourselves to the cosmological horizon in a pure de Sitter spacetime in four dimensions ($n=4$). A suitable unimodular de Sitter metric reads
\begin{equation}
\text{d}s^2=\sqrt{\omega}\left[-\left(1-\frac{\Lambda}{3}r^2\right)\text{d}t^2+\frac{1}{r^4\left(1-\frac{\Lambda}{3}r^2\right)}\text{d}R^2+r^2\left(\frac{\text{d}x^2}{1-x^2}+\left(1-x^2\right)\text{d}\phi^2\right)\right],
\end{equation}
where the coordinates are defined in the same way as for the unimodular Schwarzschild-anti-de Sitter metric presented above. The cosmological horizon is a Killing horizon with respect to the time translational Killing vector, $t^{\mu}=(1,0,0,0)$, which is timelike inside the cosmological horizon and becomes spacelike outside of it.

Consider a perturbation of the metric that satisfies the vacuum WTG equations of motion. We study the perturbation of the Hamiltonian corresponding to evolution along $t^{\mu}$ defined on a Cauchy surface $\mathcal{C}$ orthogonal to $t^{\mu}$ whose outer boundary is the horizon. Since $t^{\mu}$ is a Killing vector, $\delta H_{t}$ vanishes and we find
\begin{equation}
0=\delta H_{t}=\sqrt{\frac{\Lambda}{3}}\int_{\partial\mathcal{C}}\delta\left[\left(\sqrt{-\mathfrak{g}}/\omega\right)^{1/2}\epsilon^{\nu\mu}\right]\text{d}\mathcal{C}_{\mu\nu}+\frac{1}{8\pi}\delta\Lambda\int_{\mathcal{C}}t^{\mu}\text{d}\mathcal{C}_{\mu}.
\end{equation}
In this simple case, we can explicitly evaluate all the integrals, obtaining the following first law of the de Sitter cosmological horizon
\begin{equation}
\sqrt{\frac{\Lambda}{3}}\left(\delta\mathcal{A}_{\partial\mathcal{C}}-\frac{1}{2}\mathcal{A}_{\partial\mathcal{C}}\delta\frac{\sqrt{-\mathfrak{g}}}{\omega}\right)+\frac{1}{8\pi}\mathcal{V}_{\mathcal{C}}\delta\Lambda=0,
\end{equation}
where $\mathcal{A}_{\partial\mathcal{C}}=12\pi/\Lambda$ and $\mathcal{V}_{\mathcal{C}}=4\sqrt{3}\pi/\Lambda^{3/2}$ are the area of $\partial\mathcal{C}$ and the volume of $\mathcal{C}$, respectively. Since the background metric is unimodular, we recover the first law of the de Sitter horizon valid in GR for $\delta\mathfrak{g}=0$. Let us note that, if we choose to invoke the quantum field theory on a curved background to identify the temperature of the de Sitter horizon, $T_{\text{dS}}=\left(1/2\pi\right)\sqrt{\Lambda/3}$, the first term has the interpretation $T_{\text{dS}}\delta S$. The entropy of the de Sitter horizon then reads
\begin{equation}
\label{dS entropy Noether}
S=3\pi/\Lambda,
\end{equation}
and agrees with the GR result.

Apart from the first law, we can also obtain the Smarr formula for de Sitter spacetime which relates the volume and the area of $\mathcal{C}$. To do so, we integrate the on-shell relation for $j^{\mu}_{t}$~\eqref{j shell} over $\mathcal{C}$
\begin{equation}
\int_{\mathcal{C}}j^{\mu}_{t}\text{d}\mathcal{C}_{\mu}=-\frac{1}{8\pi}\Lambda\int_{\mathcal{C}}t^{\mu}\text{d}\mathcal{C}_{\mu}+\int_{\partial\mathcal{C}}Q_{t}^{\nu\mu}\text{d}\mathcal{C}_{\mu\nu},
\end{equation}
and use that $j^{\mu}_{t}=-Lt^{\mu}=-\Lambda t^{\mu}/4\pi$ in this case, since the symplectic potential corresponding to $t^{\mu}$ (or, in general, to any Killing vector) vanishes. The final result then reads
\begin{equation}
\frac{\Lambda \mathcal{V}_{\mathcal{C}}}{8\pi}=\frac{1}{2\pi}\sqrt{\frac{\Lambda}{3}}\frac{\mathcal{A}_{\partial\mathcal{C}}}{4},
\end{equation}
which one can easily verify by plugging in the expressions for $\mathcal{V}_{\mathcal{C}}$ and $\mathcal{A}_{\partial\mathcal{C}}$.

To conclude our discussion of de Sitter spacetime, we address an apparent inconsistency between WTG and GR pointed out in the literature~\cite{Fiol:2010} (while the issue was originally discussed in the context of unimodular gravity, it also appears in WTG). It concerns the entropy of the de Sitter horizon defined using the path integral approach. This method uses a path integral in an Euclidean space to obtain the partition function, $Z$~\cite{Gibbons:1977}. By standard thermodynamics arguments, it then holds
\begin{equation}
\label{ln Z}
\ln Z=-E/T+S,
\end{equation}
where $T$ denotes the temperature, $E$ the total energy, and $S$ the entropy. The equation may also contain additional terms corresponding to nonvanishing chemical potentials. In the case of the $4$-dimensional de Sitter spacetime, $\ln Z$ equals simply minus an integral of the Euclidean action over a $4$-sphere $V$ of radius $\sqrt{3/\Lambda}$~\cite{Fiol:2010}. In GR, it yields
\begin{equation}
I=-\frac{2\Lambda}{16\pi}\mathcal{V}_{V}=-\frac{3\pi}{\Lambda},
\end{equation}
whereas for the action of WTG we find
\begin{equation}
I=-\frac{4\Lambda-2\lambda}{16\pi}\mathcal{V}_{V}=-\frac{3\pi}{\Lambda}\left(2-\frac{\lambda}{\Lambda}\right),
\end{equation}
where $\lambda$ is a constant present in the Lagrangian, which we set to zero in the majority of the paper. As we argued in section~\ref{WTG}, $\lambda$ is unrelated to the cosmological constant and its value does not affect the dynamics of WTG in any way.

De Sitter spacetime is completely described by the value of $\Lambda$. In GR, $\Lambda$ is a fixed parameter of the Lagrangian, which we cannot change. Hence, the standard treatment of the de Sitter spacetime sets its energy to zero and considers no chemical potentials. Then, equation~\eqref{ln Z} implies $I=S$, and the entropy of the de Sitter horizon obeys $S=3\pi/\Lambda$ in GR~\cite{Gibbons:1977}. In contrast, setting $I=S$ for de Sitter spacetime in WTG yields
\begin{equation}
\label{dS entropy}
S=\frac{3\pi}{\Lambda}\left(2-\frac{\lambda}{\Lambda}\right),
\end{equation}
which agrees with the GR case only for $\Lambda=\lambda$. However, fixing $\Lambda=\lambda$ runs contrary to $\Lambda$ being an integration constant independent of the Lagrangian parameters (see the discussion at the end of subsection~\ref{current} for details of the reasoning). Equation~\eqref{dS entropy} also disagrees with the entropy of de Sitter horizon in WTG~\eqref{dS entropy Noether} we found using our Noether charge formalism.

To reconcile this discrepancy, we first note that $\Lambda$ is allowed to vary in WTG. In the previous subsection, we reviewed the arguments which lead to identifying a varying (negative) cosmological constant with pressure, $p_{\Lambda}=-\Lambda/8\pi$. If we adopt the same interpretation for a varying positive cosmological constant in de Sitter spacetime, we have by a standard thermodynamic argument
\begin{equation}
I=-S+p_{\Lambda}\mathcal{V}_{V}=-S-\frac{3\pi}{\Lambda},
\end{equation}
and, thus,
\begin{equation}
S=\frac{3\pi}{\Lambda}\left(1-\frac{\lambda}{\Lambda}\right).
\end{equation}
Regarding the contribution from $\lambda$, it can be easily interpreted in WTG. The term in WTG action proportional to $\lambda$ equals $\mathcal{V}_{V}^{\left(\omega\right)}\lambda$, where $\mathcal{V}_{V}^{\left(\omega\right)}$ is the spacetime volume of the integration domain $V$, measured with respect to the non-dynamical volume form, $\boldsymbol{\omega}$. Therefore, $\mathcal{V}_{V}^{\left(\omega\right)}\lambda$ is a universal constant which simply quantifies our freedom to perform a constant shift of the value of entropy. Hence, we can just set $\lambda=0$ in the same way it is customary to set $S=0$ at the absolute zero temperature in standard thermodynamics. In this way, the results for entropy of de Sitter horizon in GR and WTG are consistent. Furthermore, we see that the Noether charge and Euclidean canonical ensemble approaches to calculating entropy for WTG lead to equivalent results (although we have shown this only for de Sitter spacetime, the cosmological constant contribution may be expected to be the only possible obstacle to the equivalence).

\section{The first law in WTG coupled with matter}
\label{sourced}

So far we have been discussing the Noether charge formalism for vacuum WTG. Adding matter sources minimally coupled to gravity is fairly simple, as the general Noether charge formalism reviewed in subsection~\ref{general} can be straightforwardly applied. We first briefly discuss the general case and then illustrate it on the example of a stationary and asymptotically flat black hole spacetime filled with a perfect fluid.

\subsection{General formalism for matter fields}

The matter symplectic potential $\theta^{\mu}_{\psi}$ and symplectic current $\Omega^{\mu}_{\psi}$ can be obtained directly from the general equations~\eqref{dL g} and~\eqref{omega g}, respectively. Regarding the local symmetry transformations, the Noether current corresponding to local Weyl symmetry vanishes. For a transverse diffeomorphism generated by a vector field $\xi^{\mu}$ it is easy to check that $\alpha^{\mu}_{\psi,\xi}=\left(\sqrt{-\mathfrak{g}}/\omega\right)^{2k/n}L_{\psi}\xi^{\mu}$ and we have from the general equation~\eqref{j},
\begin{equation}
\label{j m}
j^{\mu}_{\psi,\xi}=\theta^{\mu}_{\psi}\left[\pounds_{\xi}\right]-L_{\psi}\xi^{\mu}.
\end{equation}
We cannot directly evaluate $j^{\mu}_{\psi,\xi}$ as we do not have an expression for $\theta^{\mu}_{\psi,\xi}$ corresponding to a general matter Lagrangian. However, the general definition of the symplectic potential~\eqref{theta} expresses its divergence in terms of the equations of motion and a variation of the Lagrangian. Hence, we can learn more about $j^{\mu}_{\psi,\xi}$ by evaluating its Weyl covariant divergence,
\begin{align}
\nonumber \tilde{\nabla}_{\mu}j^{\mu}_{\psi,\xi}=&\tilde{\nabla}_{\mu}\theta^{\mu}_{\psi,\xi}-\pounds_{\xi}\left[\left(\sqrt{-\mathfrak{g}}/\omega\right)^{2k/n}L_{\psi}\right] \\
=&-A_{\psi}\pounds_{\xi}\psi-\left(\sqrt{-\mathfrak{g}}/\omega\right)^{2k/n}\left(T^{\mu\nu}-\frac{1}{n}Tg^{\mu\nu}\right)2\nabla_{(\mu}\xi_{\nu)},
\end{align}
where $A_{\psi}=0$ are the matter equations of motion. A series of straightforward manipulations yields~\cite{Iyer:1996}
\begin{equation}
\tilde{\nabla}_{\mu}j^{\mu}_{\psi,\xi}=\tilde{\nabla}_{\mu}\left[-\left(\psi\cdot A_{\psi}\cdot\xi\right)^{\mu}-\left(\left(\sqrt{-\mathfrak{g}}/\omega\right)^{2k/n}T_{\nu}^{\:\;\mu}-\mathcal{J}\delta_{\nu}^{\mu}\right)\xi^{\nu}\right],
\end{equation}
where $\mathcal{J}$ is defined by equation~\eqref{div T} and $\left(\psi\cdot A_{\psi}\cdot\xi\right)^{\mu}$ is to be understood as $\psi A_{\psi}\xi^{\mu}$ for scalar fields and $\psi_{\nu}A_{\psi}^{(\mu}\xi^{\nu)}$ for vector fields (more general tensorial fields require a separate treatment). This form of the divergence of $j^{\mu}_{\psi,\xi}$ implies~\cite{Wald:1990b,Iyer:1996}
\begin{equation}
\label{j m 2}
j^{\mu}_{\psi,\xi}=-\left(\psi\cdot A_{\psi}\cdot\xi\right)^{\mu}-\left[\left(\sqrt{-\mathfrak{g}}/\omega\right)^{2k/n}T_{\nu}^{\:\;\mu}-\mathcal{J}\delta_{\nu}^{\mu}\right]\xi^{\nu}+\tilde{\nabla}_{\nu}Q_{\psi,\xi}^{\nu\mu},
\end{equation}
where $Q^{\nu\mu}_{\psi,\xi}$ is the antisymmetric matter Noether charge tensor. Once the matter Lagrangian is specified, the precise form of $Q_{\psi,\xi}^{\nu\mu}$ can be found by comparing equations~\eqref{j m} and~\eqref{j m 2} for $j^{\mu}_{\psi,\xi}$. Since the second term in $j^{\mu}_{\psi,\xi}$ corresponds to the right hand side of the WTG equations of motion, the combined matter and gravitational WTG Noether current on shell reduces to the divergence of the total Noether charge tensor and a contribution proportional to $\Lambda$ discussed in subsection~\ref{current}.

As an aside, we note that the matter Noether charge we derived is proportional to $\xi^{\mu}$. This can be seen from the fact that a Lagrangian $L_{\psi}$ for minimally coupled matter fields contains at most first derivatives of the matter variables (otherwise, $L_{\psi}$ would depend on the connection which would be in conflict with the assumption of minimal coupling). Hence, from the general definition of the symplectic potential~\eqref{theta}, we see that $\theta^{\mu}_{\psi}$ can depend only on variations of the matter variables and not on their derivatives. The matter field variations corresponding to transverse diffeomorphisms generated by a vector field $\xi^{\mu}$ are given by Lie derivatives along $\xi^{\mu}$, which depend on the first (but not higher) derivatives of $\xi^{\mu}$. Hence, the Noether current depends at most on the first derivatives of $\xi^{\mu}$ (see equation~\eqref{j m}). The Noether charge $Q^{\nu\mu}_{\psi,\xi}$ appears as a total divergence in the expression for $j^{\mu}_{\psi,\xi}$. Therefore, it can contain only $\xi^{\mu}$ and not its derivatives and there exists a WTDiff invariant antisymmetric tensor $W_{\rho}^{\;\:\nu\mu}=W_{\rho}^{\;\:[\nu\mu]}$, such that $Q^{\nu\mu}_{\psi,\xi}=\xi^{\rho}W_{\rho}^{\;\:\nu\mu}$.

The symplectic current corresponding to a transverse diffeomorphism and an arbitrary perturbation of the metric and matter fields equals
\begin{align}
\nonumber \Omega_{\psi}^{\mu}\left[\pounds_{\xi},\delta\right]=&\delta\theta_{\psi}^{\mu}\left[\pounds_{\xi}\right]-\pounds_{\xi}\theta_{\psi}^{\mu}\left[\delta\right]. \\
=&\delta\left(j_{\psi,\xi}^{\mu}+\left(\sqrt{-\mathfrak{g}}/\omega\right)^{2k/n}L_{\psi}\xi^{\mu}\right)-\pounds_{\xi}\theta_{\psi}^{\mu}\left[\delta\right].
\end{align}
If both the background and the perturbation satisfy the equations of motion, we have
\begin{align}
\label{omega m}
\nonumber \Omega_{\psi}^{\mu}\left[\pounds_{\xi},\delta\right]=&\tilde{\nabla}_{\nu}\left(\delta Q_{\psi,\xi}^{\nu\mu}-2\xi^{[\nu}\theta_{\psi}^{\mu]}\left[\delta\right]\right)-\delta\left[\left(\sqrt{-\mathfrak{g}}/\omega\right)^{2k/n}T_{\nu}^{\;\:\mu}-\mathcal{J}\delta^{\mu}_{\nu}\right]\xi^{\nu} \\
&+\frac{1}{2}\left(\sqrt{-\mathfrak{g}}/\omega\right)^{2k/n}\xi^{\mu}\left(T^{\alpha\beta}-\frac{1}{n}Tg^{\alpha\beta}\right)\delta g_{\alpha\beta}.
\end{align}
Finally, integrating $\Omega_{\psi}^{\mu}\left[\pounds_{\xi},\delta\right]$ over a Cauchy surface $\mathcal{C}$ yields the perturbation of the matter Hamiltonian, $H_{\psi,\xi}$,
\begin{align}
\label{H m}
\nonumber \delta H_{\psi,\xi}=&\int_{\partial\mathcal{C}}\left(\delta Q_{\psi,\xi}^{\nu\mu}-2\xi^{\nu}\theta_{\psi}^{\mu}\left[\delta\right]\right)\text{d}\mathcal{C}_{\mu\nu}-\int_{\mathcal{C}}\delta\left[\left(\sqrt{-\mathfrak{g}}/\omega\right)^{2k/n}T_{\nu}^{\;\:\mu}-\mathcal{J}\delta^{\mu}_{\nu}\right]\xi^{\nu}\text{d}\mathcal{C}_{\mu} \\
&+\frac{1}{2}\int_{\mathcal{C}}\left(\sqrt{-\mathfrak{g}}/\omega\right)^{2k/n}\left(T^{\alpha\beta}-\frac{1}{n}Tg^{\alpha\beta}\right)\delta g_{\alpha\beta}\xi^{\mu}\text{d}\mathcal{C}_{\mu}.
\end{align}
The total Hamiltonian perturbation consists of the matter part and the gravitational contribution $\delta H_{\text{g},\xi}$ (we use the subscript $\text{g}$ to distinguish it from the perturbation of the total Hamiltonian) discussed in subsection~\ref{hamiltonian}, so that $\delta H_{\xi}=\delta H_{\text{g},\xi}+\delta H_{\psi,\xi}$. The perturbation of the  matter Hamiltonian, $\delta H_{\psi,\xi}$ is given by equation~\eqref{H m}. However, equation~\eqref{dH1} for the perturbation of the gravitational Hamiltonian was derived by invoking the vacuum WTG equations of motion~\eqref{vacuum EoMs} and it no longer holds in the presence of matter. We only have
\begin{equation}
\delta H_{\text{g},\xi}=\int_{\mathcal{C}}\Omega^{\mu}_{\text{g}}\left[\pounds_{\xi},\delta\right]\text{d}\mathcal{C}_{\mu}.
\end{equation}
To find a suitable expression for the total Hamiltonian perturbation, $\delta H_{\xi}$, we can follow the same line of reasoning as we did to obtain equation~\eqref{dH1} for the vacuum case. The total symplectic current $\Omega^{\mu}=\Omega_{\text{g}}^{\mu}+\Omega_{\psi}^{\mu}$ obeys
\begin{equation}
\Omega^{\mu}\left[\pounds_{\xi},\delta\right]=\delta\theta^{\mu}\left[\pounds_{\xi}\right]-\pounds_{\xi}\theta^{\mu}\left[\delta\right],
\end{equation}
where $\theta^{\mu}=\theta^{\mu}_{\text{g}}+\theta^{\mu}_{\psi}$ denotes the total symplectic potential. This can be rewritten in the same way as for the vacuum WTG symplectic current (equations~\eqref{lie theta}-\eqref{omega 2} hold, just with the matter contributions added) and we obtain
\begin{equation}
\Omega^{\mu}\left[\pounds_{\xi},\delta\right]=\tilde{\nabla}_{\nu}\left(\delta Q^{\nu\mu}_{\xi}-2\xi^{[\nu}\theta^{\mu]}\left[\delta\right]\right)-\frac{1}{8\pi}\delta\Lambda\xi^{\mu},
\end{equation}
where $Q^{\nu\mu}_{\xi}=Q^{\nu\mu}_{\text{g},\xi}+Q^{\nu\mu}_{\psi,\xi}$ denotes the total Noether charge. The integral over a Cauchy surface $\mathcal{C}$ then yields the perturbation of the total Hamiltonian
\begin{equation}
\delta H_{\xi}=\delta H_{\text{g},\xi}+\delta H_{\psi,\xi}=\int_{\partial\mathcal{C}}\left(\delta Q^{\nu\mu}_{\xi}-2\xi^{\nu}\theta^{\mu}\left[\delta\right]\right)\text{d}\mathcal{C}_{\mu\nu}-\int_{\mathcal{C}}\frac{1}{8\pi}\delta\Lambda\xi^{\mu}\text{d}\mathcal{C}_{\mu}.
\end{equation}
Substituting for $\delta H_{\psi,\xi}=\int_{\mathcal{C}}\Omega^{\mu}_{\psi}\text{d}\mathcal{C}_{\mu}$ from equation~\eqref{omega m}, we finally obtain
\begin{align}
\nonumber \delta H_{\text{g},\xi}=\int_{\mathcal{C}}\Omega_{\text{g}}^{\mu}\left[\pounds_{\xi},\delta\right]\text{d}\mathcal{C}_{\mu}=&\int_{\partial\mathcal{C}}\left(\delta Q^{\nu\mu}_{\text{g},\xi}-2\xi^{\nu}\theta^{\mu}_{\text{g}}\left[\delta\right]\right)\text{d}\mathcal{C}_{\mu\nu}-\int_{\mathcal{C}}\frac{1}{8\pi}\delta\Lambda\xi^{\mu}\text{d}\mathcal{C}_{\mu} \\
\nonumber &-\int_{\mathcal{C}}\delta\left[\left(\sqrt{-\mathfrak{g}}/\omega\right)^{2k/n}T_{\nu}^{\;\:\mu}-\mathcal{J}\delta^{\mu}_{\nu}\right]\xi^{\nu}\text{d}\mathcal{C}_{\mu} \\
&+\frac{1}{2}\int_{\mathcal{C}}\left(\sqrt{-\mathfrak{g}}/\omega\right)^{2k/n}\left(T^{\alpha\beta}-\frac{1}{n}Tg^{\alpha\beta}\right)\delta g_{\alpha\beta}\xi^{\mu}\text{d}\mathcal{C}_{\mu}, \label{dH m+g}
\end{align}
where the contributions from $\delta Q_{\psi,\xi}^{\nu\mu}$ and $\theta^{\mu}_{\psi}\left[\delta\right]$ cancel out. Equation~\eqref{dH m+g} allows us to straightforwardly derive the first law of black hole mechanics for a stationary black hole spacetime with matter fields present, as we show in detail in the following.

Consider a stationary, asymptotically flat black hole spacetime with arbitrary minimally coupled matter fields present. As in the vacuum case we discussed in subsection~\ref{black holes} the spacetime has a time translational Killing vector, $t^{\mu}$, and $n-3$ rotational Killing vectors, $\varphi_{(i)}^{\mu}$. The black hole horizon is again a Killing horizon with respect to the Killing vector $\xi^{\mu}=t^{\mu}+\sum_{i=1}^{n-3}\Omega_{\mathcal{H}}^{(i)}\varphi_{(i)}^{\mu}$, where $\Omega_{\mathcal{H}}^{(i)}$ are the constant angular velocities of the horizon. The spacetime also possesses a spacelike Cauchy surface, $\mathcal{C}$, whose boundary is composed of its intersections with the spatial infinity, $\mathcal{C}_{\infty}$, and with the horizon, $\partial\mathcal{C}_{\mathcal{H}}$.

Although all the physical quantities describing the matter are required to have the same symmetries as the spacetime, the matter Lagrangian may also contain dynamical variables which are not stationary~\cite{Iyer:1996} (in other words, $\pounds_{t}\psi\ne0$, $\pounds_{\varphi_{(i)}}\psi\ne0$). A typical example of this behaviour are perfect fluids, whose Lagrangian depends on Lagrange multipliers which in general do not share the symmetries of the spacetime, although fluid's entropy, temperature, velocity, particle density, energy density and pressure do. We demand that there exists a vector field co-moving with the matter that is of the form $U^{\mu}=t^{\mu}+\sum_{i=1}^{n-3}\Omega^{(i)}\varphi_{(i)}^{\mu}$, where $\Omega^{(i)}$ are the matter angular velocities (not necessarily constant) in the various Killing directions.

Let us introduce an arbitrary perturbation of the metric and the fluid variables which obeys the equations of motion and preserves the asymptotic flatness of the spacetime. We are interested in evaluating the perturbation of the gravitational Hamiltonian, $\delta H_{\text{g},\xi}$, corresponding to the evolution along the Killing vector $\xi^{\mu}$. Since $\xi^{\mu}$ is a Killing vector, \mbox{$\pounds_{\xi}\tilde{g}_{\mu\nu}=0$}, the gravitational symplectic current $\Omega^{\mu}\left[\pounds_{\xi},\delta\right]$ vanishes and so does $\delta H_{\text{g},\xi}$. Applying equation~\eqref{dH m+g} for the perturbation of the total Hamiltonian to this case then yields
\begin{align}
\nonumber &\int_{\partial\mathcal{C}}\left(\delta Q^{\nu\mu}_{\text{g},\xi}-2\xi^{\nu}\theta^{\mu}_{\text{g}}\left[\delta\right]\right)\text{d}\mathcal{C}_{\mu\nu}-\int_{\mathcal{C}}\delta\left[\left(\sqrt{-\mathfrak{g}}/\omega\right)^{2k/n}T_{\nu}^{\;\:\mu}-\mathcal{J}\delta^{\mu}_{\nu}\right]\xi^{\nu}\text{d}\mathcal{C}_{\mu} \\
&+\frac{1}{2}\int_{\mathcal{C}}\left(\sqrt{-\mathfrak{g}}/\omega\right)^{2k/n}\left(T^{\alpha\beta}-\frac{1}{n}Tg^{\alpha\beta}\right)\delta g_{\alpha\beta}\xi^{\mu}\text{d}\mathcal{C}_{\mu}=0. \label{dH matter}
\end{align}
The surface integral comes from the gravitational degrees of freedom and has the same interpretation as in the vacuum case:
\begin{equation}
\int_{\partial\mathcal{C}}\left(\delta Q^{\nu\mu}_{\text{g},\xi}-2\xi^{\nu}\theta^{\mu}_{\text{g}}\left[\delta\right]\right)\text{d}\mathcal{C}_{\mu\nu}=\delta E-\sum_{i=1}^{n-3}\Omega^{(i)}_{\mathcal{H}}\delta J_{(i)}-\frac{1}{8\pi}\kappa\int_{\partial\mathcal{C}_{\mathcal{H}}}\delta\left[\left(\sqrt{-\mathfrak{g}}/\omega\right)^{2/n}\epsilon^{\nu\mu}\right]\text{d}\mathcal{C}_{\mu\nu}. \label{BH part}
\end{equation}
In this case the perturbations of the total energy and angular momenta include both the contributions of the black hole and the matter fields. The first volume integral in equation~\eqref{dH matter} can be rewritten in the following way
\begin{align}
\nonumber &-\int_{\mathcal{C}}\delta\left[\left(\sqrt{-\mathfrak{g}}/\omega\right)^{2k/n}T_{\nu}^{\;\:\mu}-\mathcal{J}\delta^{\mu}_{\nu}\right]\xi^{\nu}\text{d}\mathcal{C}_{\mu}=-\int_{\mathcal{C}}\delta\left[\left(\sqrt{-\mathfrak{g}}/\omega\right)^{2k/n}T_{\nu}^{\;\:\mu}\right]U^{\nu}\text{d}\mathcal{C}_{\mu} \\
&-\int_{\mathcal{C}}\delta\mathcal{J}\xi^{\mu}\text{d}\mathcal{C}_{\mu}+\int_{\mathcal{C}}\sum_{i=1}^{n-3}\left(\Omega^{(i)}-\Omega^{(i)}_{\mathcal{H}}\right)\delta\left[\left(\sqrt{-\mathfrak{g}}/\omega\right)^{2k/n}T_{\nu}^{\;\:\mu}\right]\varphi_{(i)}^{\nu}\text{d}\mathcal{C}_{\mu}.
\end{align}
The first term now contains a variation of the energy-momentum tensor contracted with the vector $U^{\nu}$ co-moving with the matter. The second term represents the contribution of the local non-conservation of energy-momentum. The last integral contains the perturbations of the WTDiff invariant angular momenta densities of the matter, $\delta\tilde{J}_{(i)}^{\mu}$, defined in the standard way
\begin{equation}
\delta\tilde{J}_{(i)}^{\mu}=\delta\left[\left(\sqrt{-\mathfrak{g}}/\omega\right)^{2k/n}T_{\nu}^{\;\:\mu}\right]\varphi_{(i)}^{\nu}.
\end{equation}
Since the total angular momenta in equation~\eqref{BH part} obey $J_{(i)}=J_{\mathcal{H}}^{(i)}+\int_{\mathcal{C}}\tilde{J}_{(i)}^{\mu}\text{d}\mathcal{C}_{\mu}$, it holds
\begin{equation}
-\sum_{i=1}^{n-3}\Omega^{(i)}_{\mathcal{H}}\delta J_{(i)}+\int_{\mathcal{C}}\sum_{i=1}^{n-3}\left(\Omega^{(i)}-\Omega^{(i)}_{\mathcal{H}}\right)\delta\tilde{J}_{(i)}^{\mu}\text{d}\mathcal{C}_{\mu}=-\sum_{i=1}^{n-3}\Omega^{(i)}_{\mathcal{H}}\delta J^{(i)}_{\mathcal{H}}+\int_{\mathcal{C}}\sum_{i=1}^{n-3}\Omega^{(i)}\delta\tilde{J}_{(i)}^{\mu}\text{d}\mathcal{C}_{\mu}.
\end{equation}
In total, equation~\eqref{dH matter} yields the following form of the first law for stationary, asymptotically flat black hole spacetimes with matter fields
\begin{align}
\nonumber &\delta E-\sum_{i=1}^{n-3}\Omega^{(i)}_{\mathcal{H}}\delta J^{(i)}_{\mathcal{H}}-\frac{1}{8\pi}\kappa\int_{\partial\mathcal{C}_{\mathcal{H}}}\delta\left[\left(\sqrt{-\mathfrak{g}}/\omega\right)^{2/n}\epsilon^{\nu\mu}\right]\text{d}\mathcal{C}_{\mu\nu} \\
\nonumber &-\int_{\mathcal{C}}\delta\left[\left(\sqrt{-\mathfrak{g}}/\omega\right)^{2k/n}T_{\nu}^{\;\:\mu}\right]U^{\nu}\text{d}\mathcal{C}_{\mu}+\frac{1}{2}\int_{\mathcal{C}}\left(\sqrt{-\mathfrak{g}}/\omega\right)^{2k/n}\left(T^{\alpha\beta}-\frac{1}{n}Tg^{\alpha\beta}\right)\delta g_{\alpha\beta}\xi^{\mu}\text{d}\mathcal{C}_{\mu} \\
&+\int_{\mathcal{C}}\sum_{i=1}^{n-3}\Omega^{(i)}\delta\tilde{J}_{(i)}^{\mu}\text{d}\mathcal{C}_{\mu}+\int_{\mathcal{C}}\delta\mathcal{J}\xi^{\mu}\text{d}\mathcal{C}_{\mu}=0. \label{dH matter 2}
\end{align}
Any further analysis of the first law depends on the specific properties of the matter fields present in the spacetime.

\subsection{WTDiff invariant perfect fluids}

The original derivation of the first law of black hole mechanics in GR considered a stationary, asymptotically flat black hole spacetime filled with a perfect fluid~\cite{Bardeen:1973}. This form of the first law was later reproduced (and generalised) using the Noether charge formalism for Diff invariant gravity~\cite{Iyer:1996}. Here, we derive it for WTG. Our aim is to both illustrate the matter field contributions to the Hamiltonian perturbation discussed above in a general setting and to show the physical equivalence of the final formula with the result known in GR.

Before applying the Noether charge formalism to a perfect fluid, we first need to introduce a suitable Lagrangian description. In particular, we develop a WTDiff invariant (and somewhat simplified) version of the formalism presented in~\cite{Brown:1993}. We choose the entropy per particle, $s$, and the particle density, $\nu$, as the configuration variables describing the fluid. We then consider an equation of state which expresses the energy density of the fluid as a function of $s$ and $\nu$, $\rho=\rho\left(s,\nu\right)$. We further need the velocity of the fluid, $u^{\mu}$, normalised so that $u_{\mu}u^{\mu}=-1$. To keep this normalisation Weyl invariant, the behaviour of $u^{\mu}$ under Weyl transformations must be $u'^{\mu}=e^{-\sigma}u^{\mu}$ (in other words, $g_{\mu\nu}+u_{\mu}u_{\nu}$ must be a projector to the subspace orthogonal to $u^{\mu}$ in every gauge). Using the fluid velocity and particle number density, we introduce the particle number density flux
\begin{equation}
\label{particle flux}
I^{\mu}=\left(\sqrt{-\mathfrak{g}}/\omega\right)^{1/n}\nu u^{\mu},
\end{equation}
which is Weyl invariant.

As the basic component of our fluid Lagrangian we choose the energy density, $\rho\left(\nu,s\right)$. Furthermore, it must incorporate the conditions that the fluxes of particle number density and entropy per particle along the flow lines are conserved (these requirements characterise a perfect fluid), which we add via Lagrange multipliers. In total, the Lagrangian reads\footnote{In principle, one should also add a term depending on the Lagrangian coordinates of the fluid, which specifies the flow lines and serves to fix them on the spacetime boundaries~\cite{Brown:1993}. However, this is not needed for our purposes.}
\begin{equation}
L_{\text{f}}=-\rho\left(s,\nu\right)+I^{\mu}\left(\tilde{\nabla}_{\mu}\eta+s\tilde{\nabla}_{\mu}\tau\right),
\end{equation}
where $\eta$ and $\tau$ are spacetime scalars which play the role of Lagrange multipliers. We discuss their physical interpretation after deriving the equations of motion. Varying the matter action with respect to the metric yields the traceless part of the energy-momentum tensor
\begin{equation}
\label{fluid emt}
T_{\mu\nu}-\frac{1}{n}Tg_{\mu\nu}=\left(\rho+p\right)\left(u_{\mu}u_{\nu}+\frac{1}{n}g_{\mu\nu}\right),
\end{equation}
where we identified the pressure
\begin{equation}
\label{pressure}
p=\nu\frac{\partial\rho}{\partial\nu}-\rho,
\end{equation}
by comparison with the standard form of the energy-momentum tensor of a perfect fluid. The variations with respect to $\eta$ and $\tau$ give us the conservation of the particle number density flux and the entropy per particle flux, respectively,
\begin{align}
\tilde{\nabla}_{\mu}I^{\mu}=0,\qquad\tilde{\nabla}_{\mu}\left(sI^{\mu}\right)=0.
\end{align}
The variation with respect to $s$ implies
\begin{equation}
\label{d/ds}
-\frac{\partial\rho}{\partial s}+I^{\mu}\tilde{\nabla}_{\mu}\tau=0,
\end{equation}
which can be interpreted as the first law of thermodynamics for the fluid~\cite{Brown:1993}. Indeed, if we specify the Weyl invariant fluid temperature
\begin{equation}
\label{temperature}
\mathcal{T}=\left(\sqrt{-\mathfrak{g}}/\omega\right)^{1/n}u^{\mu}\tilde{\nabla}_{\mu}\tau,
\end{equation}
equation~\eqref{d/ds} becomes
\begin{equation}
\label{T}
\mathcal{T}=\frac{1}{\nu}\frac{\partial\rho}{\partial s}.
\end{equation}
Next, varying the fluid action with respect to $\nu$ leads to
\begin{equation}
\label{delta nu}
\frac{\partial \rho}{\partial\nu}-\left(\sqrt{-\mathfrak{g}}/\omega\right)^{1/n}u^{\mu}\left(\tilde{\nabla}_{\mu}\eta+s\tilde{\nabla}_{\mu}\tau\right)=0.
\end{equation}
The last term equals $-\mathcal{T}s$ according to equation~\eqref{temperature} and the first term corresponds to $\left(\rho+p\right)/\nu$ (see equation~\eqref{pressure}). Using the Gibbs-Duhem equation of standard thermodynamics, we obtain the chemical potential $\mu$ of the fluid
\begin{equation}
\label{mu}
\mu=\frac{\rho+p}{\nu}-\mathcal{T}s=\frac{\partial \rho}{\partial\nu}-s\left(\sqrt{-\mathfrak{g}}/\omega\right)^{1/n}u^{\mu}\tilde{\nabla}_{\mu}\tau.
\end{equation}
Equation~\eqref{delta nu} then relates the chemical potential with the Lagrange multiplier $\eta$
\begin{equation}
\mu=\left(\sqrt{-\mathfrak{g}}/\omega\right)^{1/n}u^{\mu}\tilde{\nabla}_{\mu}\eta.
\end{equation}
Lastly, a variation of the fluid action with respect to $u^{\mu}$ yields an equation which governs the behaviour of $\eta$ and $\tau$ on the surfaces orthogonal to the flow lines.

It is not difficult to see that the fluid equations of motion imply that the energy-momentum tensor of the perfect fluid is divergenceless, $\tilde{\nabla}_{\nu}T_{\mu}^{\;\:\nu}=0$~\cite{Brown:1993}. Hence, local energy-momentum conservation is directly built into our construction (in other words, \mbox{$\mathcal{J}=0$} in our case).

We now apply the Noether charge formalism for matter fields that we described in the previous subsection to our perfect fluid Lagrangian. We straightforwardly obtain the symplectic potential for a general variation of the fluid variables and the metric, and the Noether current corresponding to a transverse diffeomorphism generated by a vector \mbox{field $\xi^{\mu}$}
\begin{equation}
\theta^{\mu}_{\text{f}}=I^{\mu}\left(\delta\eta+s\delta\tau\right),\qquad\;j^{\mu}_{\text{f},\xi}=-T_{\nu}^{\;\:\mu}\xi^{\nu}. \label{theta f}
\end{equation}
The Noether charge $Q^{\nu\mu}_{\text{f},\xi}$ identically vanishes in this case.

Now consider some solution of the WTG equations of motion with the perfect fluid energy-momentum tensor. For the symplectic current $\Omega^{\mu}_{\text{f}}\left[\pounds_{\xi},\delta\right]$ corresponding to a transverse diffeomorphism and an arbitrary perturbation of the fluid variables and the metric which satisfies the equations of motion, equation~\eqref{omega m} implies
\begin{equation}
\Omega_{\text{f}}^{\mu}=-2\tilde{\nabla}_{\nu}\left(\xi^{[\nu}\theta_{\text{f}}^{\mu]}\right)-\xi^{\nu}\delta T_{\nu}^{\;\:\mu}+\frac{1}{2}\xi^{\mu}\left(T^{\alpha\beta}-\frac{1}{n}Tg^{\alpha\beta}\right)\delta g_{\alpha\beta},
\end{equation}
which directly yields the perturbation of the fluid Hamiltonian for the evolution along $\xi^{\mu}$, $\delta H_{\xi,\text{f}}$ (see equation~\eqref{H m}).

\subsection{WTG first law of black hole mechanics with a perfect fluid}

Upon introducing a suitable description of a perfect fluid, we return to deriving the first law for a stationary, asymptotically flat black hole spacetime filled with a perfect fluid. Our starting point is equation~\eqref{dH matter 2} valid for general matter fields. The perturbation of the fluid energy-momentum tensor satisfies
\begin{equation}
U^{\nu}\delta T_{\nu}^{\;\:\mu}=U^{\nu}\delta\left[\left(\rho+p\right)u_{\nu}u^{\mu}+p\delta^{\mu}_{\nu}\right]=U^{\nu}\delta\left[\left(\nu\mu+\nu\mathcal{T}s\right)u_{\nu}u^{\mu}\right]+U^{\mu}\delta p,
\end{equation}
where the fluid velocity is given by $u^{\mu}=U^{\mu}/\sqrt{-g_{\alpha\beta}U^{\alpha}U^{\beta}}=U^{\mu}/\vert U\vert$ and, since $\delta U^{\mu}=0$, we can use that $U^{\nu}\delta u_{\nu}=\vert U\vert u^{\alpha}u^{\beta}\delta g_{\alpha\beta}/2$. Furthermore, equations~\eqref{pressure},~\eqref{T} and~\eqref{delta nu} together imply $\delta p+\nu\delta\mu+\nu s\delta\mathcal{T}=0$. We further use the definition of the particle number density flux $I^{\mu}$~\eqref{particle flux}. The final result after some calculations is
\begin{align}
U^{\nu}\delta T_{\nu}^{\;\:\mu}=&\frac{1}{2}U^{\nu}\left(\rho+p\right)\left(u_{\nu}u^{\mu}+\frac{1}{n}\delta_{\mu}^{\nu}\right)+\left(\sqrt{-\mathfrak{g}}/\omega\right)^{-1/n}\vert U\vert\mu\delta I^{\mu} \\
&+\left(\sqrt{-\mathfrak{g}}/\omega\right)^{-1/n}\vert U\vert\mathcal{T}\delta\left[\left(\sqrt{-\mathfrak{g}}/\omega\right)^{1/n}\nu su^{\mu}\right].
\end{align}
The first term is just the traceless part of the energy-momentum tensor and it cancels out with the last integral in equation~\eqref{dH matter 2}. The final form of the first law of black hole mechanics for a stationary, asymptotically flat black hole spacetime filled with a perfect fluid thus reads
\begin{align}
\nonumber 0=&\delta E-\Omega^{(i)}_{\mathcal{H}}\delta J_{(i)}-\frac{1}{8\pi}\kappa\int_{\partial\mathcal{C}_{\mathcal{H}}}\delta\left[\left(\sqrt{-\mathfrak{g}}/\omega\right)^{2/n}\epsilon^{\nu\mu}\right]\text{d}\mathcal{C}_{\mu\nu} \\
&-\int_{\mathcal{C}}\left(\sqrt{-\mathfrak{g}}/\omega\right)^{-1/n}\vert U\vert\mu\delta I^{\mu}\text{d}\mathcal{C}_{\mu}-\int_{\mathcal{C}}\left(\sqrt{-\mathfrak{g}}/\omega\right)^{-1/n}\vert U\vert\mathcal{T}\delta\tilde{S}^{\mu}\text{d}\mathcal{C}_{\mu}-\int_{\mathcal{C}}\sum_{i=1}^{n-3}\Omega^{(i)}\delta\tilde{J}^{\mu}_{(i)}\text{d}\mathcal{C}_{\mu},
\end{align}
where the first term of the second line quantifies the change of the fluid energy due to absorption of particles by the black hole, as it contains the perturbation of the previously defined particle number density current, $I^{\mu}$, and the chemical potential, $\mu$, with the red-shift between the horizon and the asymptotic infinity taken into account by the factor $\left(\sqrt{-\mathfrak{g}}/\omega\right)^{-1/n}\vert U\vert$. The second term of the second line corresponds to heat exchange between the fluid and the black hole, as $\tilde{S}^{\mu}=\left(\sqrt{-\mathfrak{g}}/\omega\right)^{1/n}su^{\mu}$ is the WTDiff invariant flux of the entropy density and $\left(\sqrt{-\mathfrak{g}}/\omega\right)^{-1/n}\vert U\vert\mathcal{T}$ the red-shifted temperature. In the unimodular gauge, and for perturbations that do not change the metric determinant, $\delta\mathfrak{g}=0$, we of course recover the GR form of the first law~\cite{Bardeen:1973,Iyer:1996}.

\section{The first law for causal diamonds}
\label{conformal}

The Noether charge formalism for Diff invariant theories of gravity has been used to derive the first law of causal diamonds, which plays an important role in thermodynamics of spacetime~\cite{Jacobson:2015,Bueno:2017,Svesko:2019,Jacobson:2019}. However, applying the WTG Noether charge formalism to them leads to technical difficulties. The reason is that causal diamonds posses an isometry generated by a conformal Killing vector. Such vectors are defined by the conformal Killing equation
\begin{equation}
\delta_{\zeta}g_{\mu\nu}=\pounds_{\zeta}g_{\mu\nu}=2\nabla_{(\mu}\zeta_{\nu)}=\frac{1}{n}\nabla_{\rho}\zeta^{\rho}g_{\mu\nu}.
\end{equation}
A conformal Killing vector $\zeta^{\mu}$ does not satisfy the transversality condition as $\tilde{\nabla}_{\mu}\zeta^{\mu}\ne0$. Hence, the Noether charge formalism we developed for transverse diffeomorphisms cannot be applied directly to this case. Nevertheless, $\zeta^{\mu}$ generates a pure Weyl transformation. Hence, it does not affect the auxiliary metric (when $\tilde{\nabla}_{\mu}\zeta^{\mu}\ne0$, one must be mindful that $\delta\omega=0$ and apply the rules for Lie deriving tensor densities)
\begin{align}
\delta_{\zeta}\tilde{g}_{\mu\nu}=&\omega^{2/n}\pounds_{\zeta}\left[\left(\sqrt{-\mathfrak{g}}\right)^{-2/n}g_{\mu\nu}\right]=2\tilde{\nabla}_{\sigma}\tilde{\nabla}_{(\nu}\left(\tilde{g}_{\rho)\lambda}\zeta^{\lambda}\right)-\frac{2}{n}\tilde{g}_{\nu\rho}\tilde{\nabla}_{\sigma}\tilde{\nabla}_{\lambda}\zeta^{\lambda}=0. 
\end{align}
The transformation generated by $\zeta^{\mu}$ thus lies in the intersection of Diff and WTDiff groups and represents a symmetry transformation of WTG. However, the condition on a vector field to generate a transformation from the intersection of Diff and WTDiff groups 
which is not a transverse diffeomorphism explicitly depends on the metric~\cite{Alvarez:2006}. This prevents us from finding a general expression for the Noether current and Noether charge in this case. Nevertheless, the symplectic potential $\theta^{\mu}\left[\delta_{\zeta}\right]$ and the symplectic current $\Omega^{\mu}\left[\delta_{\zeta},\delta\right]$ corresponding to the transformation generated by $\zeta^{\mu}$ obey equations~\eqref{theta} and~\eqref{omega}, respectively, which we derived for completely general variations of the metric. Therefore, if the Hamiltonian corresponding to the evolution along $\zeta^{\mu}$ exists, its on-shell perturbation is given by the standard expression~\eqref{hamilton eq}
\begin{equation}
\delta H_{\zeta}=\Omega\left[\delta_{\zeta},\delta\right]=\int_{\mathcal{C}}\Omega^{\mu}\left[\delta_{\zeta},\delta\right]\text{d}\mathcal{C}_{\mu},
\end{equation}
where $\mathcal{C}$ is a Cauchy surface. This equation is then sufficient to derive the first law of causal diamonds in the same way we derived the first law of black hole mechanics in subsection~\ref{black holes}.

We first provide an expression for $\delta H_{\zeta}$ valid for any vector field $\zeta^{\mu}$. Then, we use it to derive the first law of causal diamonds and show that it is physically equivalent to the one valid in GR. In the present work, we concentrate on the vacuum case. The matter contribution to the first law of causal diamonds will be discussed elsewhere.

\subsection{Hamiltonian}

In the following, we consider an arbitrary vector $\zeta^{\mu}$, which need not obey the transversality condition. Our aim is to derive a perturbation of the Hamiltonian for evolution along $\zeta^{\mu}$. Since $\zeta^{\mu}$ does not in general generate a symmetry transformation of WTG, we cannot follow the procedure based on the Noether current considered in section~\ref{Noether}. Instead, we take advantage of having an explicit expression for the WTG Lagrangian and obtain the Hamiltonian perturbation by directly evaluating the symplectic current
\begin{equation}
\label{omega z}
\Omega^{\mu}\left[\delta_{\zeta},\delta\right]=\delta\theta^{\mu}\left[\delta_{\zeta}\right]-\delta_{\zeta}\theta^{\mu}\left[\delta\right],
\end{equation}
where we assume that both the background metric and the perturbation obey the vacuum WTG equations of motion.

The first term in equation~\eqref{omega z} is a perturbation of the symplectic potential $\theta^{\mu}\left[\delta_{\zeta}\right]$, which can be expressed directly from the general equation~\eqref{theta} as
\begin{equation}
\theta^{\mu}\left[\delta_{\zeta}\right]=\frac{1}{16\pi}\left(\frac{\sqrt{-\mathfrak{g}}}{\omega}\right)^{\frac{4}{n}}\left(g^{\mu\nu}g^{\rho\sigma}-g^{\mu\sigma}g^{\nu\rho}\right)\tilde{\nabla}_{\sigma}\delta_{\zeta}\tilde{g}_{\nu\rho},
\end{equation}
where
\begin{equation}
\tilde{\nabla}_{\sigma}\delta_{\zeta}\tilde{g}_{\nu\rho}=2\tilde{\nabla}_{\sigma}\tilde{\nabla}_{(\nu}\left(\tilde{g}_{\rho)\lambda}\zeta^{\lambda}\right)-\frac{2}{n}\tilde{g}_{\nu\rho}\tilde{\nabla}_{\sigma}\tilde{\nabla}_{\lambda}\zeta^{\lambda}.
\end{equation}
A straightforward calculation leads to
\begin{equation}
\label{theta z}
\theta^{\mu}\left[\delta_{\zeta}\right]=\frac{1}{8\pi}\tilde{g}^{\mu\rho}\tilde{R}_{\rho\nu}\zeta^{\nu}+\tilde{\nabla}_{\nu}\left(\frac{1}{8\pi}\left(\sqrt{-g}/\omega\right)^{2/n}\tilde{\nabla}^{[\nu}\zeta^{\mu]}\right)+\Pi^{\mu}_{\zeta},
\end{equation}
where we write
\begin{equation}
\label{pi}
\Pi^{\mu}_{\zeta}=\frac{1}{8\pi}\frac{n-1}{n}\tilde{g}^{\mu\nu}\tilde{\nabla}_{\nu}\tilde{\nabla}_{\rho}\zeta^{\rho},
\end{equation}
for the only extra term appearing with respect to the transverse diffeomorphisms case. It can be noticed that the first term on the right hand side of equation~\eqref{theta z} has formally the same form as a Weyl covariant divergence of the WTG Noether charge corresponding to $\zeta^{\mu}$. However, since $\zeta^{\mu}$ does not in general generate a symmetry transformation of WTG, we cannot understand this term as a Noether charge.

Therefore, for the on-shell perturbation of the symplectic potential, $\delta\theta^{\mu}\left[\delta_{\zeta}\right]$, we find, invoking the vacuum WTG equations of motion,
\begin{equation}
\delta\theta^{\mu}\left[\delta_{\zeta}\right]=\frac{1}{8\pi}\frac{1}{n}\zeta^{\mu}\delta\tilde{R}+\frac{1}{8\pi}\tilde{\nabla}_{\nu}\left[\delta\left(\left(\sqrt{-\mathfrak{g}}/\omega\right)^{2/n}\tilde{\nabla}^{[\nu}\zeta^{\mu]}\right)\right]+\delta\Pi^{\mu}_{\zeta}.
\end{equation}

For the second term in equation~\eqref{omega z}, $\delta_{\zeta}\theta^{\mu}\left[\delta\right]$, we find from the definition of the Lie derivative (one must be mindful that $\theta^{\mu}\left[\delta\right]$ contains the metric determinant and its derivatives, which are not tensors)
\begin{equation}
\delta_{\zeta}\theta^{\mu}\left[\delta\right]=\zeta^{\nu}\tilde{\nabla}_{\nu}\theta^{\mu}\left[\delta\right]-\theta^{\nu}\left[\delta\right]\tilde{\nabla}_{\nu}\zeta^{\mu}+\frac{2}{n}\theta^{\mu}\left[\delta\right]\tilde{\nabla}_{\nu}\zeta^{\nu}+\frac{1}{16\pi}\tilde{\nabla}_{\nu}\tilde{\nabla}_{\rho}\zeta^{\rho}\delta\tilde{g}^{\mu\nu}.
\end{equation}

In total, the symplectic current reads
\begin{align}
\nonumber \Omega^{\mu}\left[\delta_{\zeta},\delta\right]=&\frac{1}{8\pi}\frac{1}{n}\zeta^{\mu}\delta\tilde{R}+\frac{1}{8\pi}\tilde{\nabla}_{\nu}\left[\delta\left(\left(\sqrt{-\mathfrak{g}}/\omega\right)^{2/n}\tilde{\nabla}^{[\nu}\zeta^{\mu]}\right)\right]-\zeta^{\nu}\tilde{\nabla}_{\nu}\theta^{\mu}\left[\delta\right]+\theta^{\nu}\left[\delta\right]\tilde{\nabla}_{\nu}\zeta^{\mu} \\
&-\frac{2}{n}\theta^{\mu}\left[\delta\right]\tilde{\nabla}_{\nu}\zeta^{\nu}+\frac{1}{16\pi}\frac{n-2}{n}\tilde{\nabla}_{\nu}\tilde{\nabla}_{\rho}\zeta^{\rho}\delta\tilde{g}^{\mu\nu}.
\end{align}
where we used the definition of $\Pi^{\mu}_{\zeta}$~\eqref{pi}. Adding and subtracting $\zeta^{\mu}\tilde{\nabla}_{\nu}\theta^{\nu}-\theta^{\mu}\tilde{\nabla}_{\nu}\zeta^{\nu}$, we obtain a total divergence term of the form $-\tilde{\nabla}_{\nu}\left(\zeta^{[\nu}\theta^{\mu]}\left[\delta\right]\right)$,
\begin{align}
\nonumber \Omega^{\mu}\left[\delta_{\zeta},\delta\right]=&\frac{1}{8\pi}\frac{1}{n}\zeta^{\mu}\delta\tilde{R}+\frac{1}{8\pi}\tilde{\nabla}_{\nu}\left[\delta\left(\left(\sqrt{-\mathfrak{g}}/\omega\right)^{2/n}\tilde{\nabla}^{[\nu}\zeta^{\mu]}\right)\right]-\tilde{\nabla}_{\nu}\left(\zeta^{[\nu}\theta^{\mu]}\left[\delta\right]\right)-\zeta^{\mu}\tilde{\nabla}_{\nu}\theta^{\nu}\left[\delta\right] \\
&+\frac{n-2}{n}\theta^{\mu}\left[\delta\right]\tilde{\nabla}_{\nu}\zeta^{\nu}-\frac{1}{16\pi}\tilde{\nabla}_{\nu}\tilde{\nabla}_{\rho}\zeta^{\rho}\delta\tilde{g}^{\mu\nu}.
\end{align}
Finally, we use the on-shell relation
\begin{equation}
\delta L=\frac{1}{16\pi}\delta\tilde{R}=\frac{1}{16\pi}\frac{2n}{n-2}\delta\Lambda=\tilde{\nabla}_{\mu}\theta^{\mu}\left[\delta\right],
\end{equation}
to write
\begin{align}
\nonumber \Omega^{\mu}\left[\delta_{\zeta},\delta\right]=&-\frac{1}{8\pi}\zeta^{\mu}\delta\Lambda+\frac{1}{8\pi}\tilde{\nabla}_{\nu}\left[\delta\left(\left(\sqrt{-\mathfrak{g}}/\omega\right)^{2/n}\tilde{\nabla}^{[\nu}\zeta^{\mu]}\right)\right]-\tilde{\nabla}_{\nu}\left(\zeta^{[\nu}\theta^{\mu]}\left[\delta\right]\right)-\zeta^{\mu}\tilde{\nabla}_{\nu}\theta^{\nu}\left[\delta\right] \\
&+\frac{n-2}{n}\theta^{\mu}\left[\delta\right]\tilde{\nabla}_{\nu}\zeta^{\nu}+\frac{1}{16\pi}\frac{n-2}{n}\tilde{\nabla}_{\nu}\tilde{\nabla}_{\rho}\zeta^{\rho}\delta\tilde{g}^{\mu\nu}.
\end{align}
An integral of $\Omega^{\mu}\left[\delta_{\zeta},\delta\right]$ over a Cauchy surface $\mathcal{C}$ then yields the perturbation of the Hamiltonian $H_{\zeta}$ corresponding to the evolution along $\zeta^{\mu}$ (if it exists)
\begin{align}
\label{dH}
\nonumber \delta H_{\zeta}=&\int_{\mathcal{C}}\Omega^{\mu}\left[\delta_{\zeta},\delta\right]\text{d}\mathcal{C}_{\mu}=\int_{\partial\mathcal{C}}\bigg\{\frac{1}{8\pi}\delta\left[\left(\sqrt{-\mathfrak{g}}/\omega\right)^{2/n}\tilde{\nabla}^{[\nu}\zeta^{\mu]}\right]-2\zeta^{\nu}\theta^{\mu}\left[\delta\right]\bigg\}\text{d}\mathcal{C}_{\mu\nu} \\
&-\int_{\mathcal{C}}\frac{1}{8\pi}\delta\Lambda\zeta^{\mu}\text{d}\mathcal{C}_{\mu}+\frac{n-2}{n}\int_{\mathcal{C}}\left(\frac{1}{16\pi}\tilde{\nabla}_{\nu}\tilde{\nabla}_{\rho}\zeta^{\rho}\delta\tilde{g}^{\mu\nu}+\theta^{\mu}\left[\delta\right]\tilde{\nabla}_{\nu}\zeta^{\nu}\right)\text{d}\mathcal{C}_{\mu}.
\end{align}
Even for $\delta\Lambda=0$, the perturbation of the Hamiltonian contains a volume integral. We discuss its interpretation on the example of a causal diamond in the following subsection.

\subsection{Causal diamonds}

Let us now apply the previously derived formalism to a simple yet interesting case of causal diamonds. A geodesic local causal diamond centred at some spacetime point, $P$, is fully described by an arbitrary unit timelike vector, $n^{\mu}\left(P\right)$, and a length scale, $l$. We define it as a region causally determined by a spacelike geodesic ball $\Sigma_0$, which is formed by geodesics of parameter length $l$ starting in $P$ and orthogonal to $n^{\mu}\left(P\right)$. In the following, we assume that the causal diamond is constructed in a flat spacetime. We choose the time coordinate so that $n\left(P\right)=\partial/\partial t$, and spatial coordinates $\lbrace x^{i}\rbrace_{i=1}^{i=n-1}$ to be Cartesian. We further perform a suitable Weyl transformation to have (at least locally) a unimodular metric, $\sqrt{-\mathfrak{g}}=\omega$. We only consider a unimodular form of the metric for computational convenience (as in subsection~\ref{lambda}). In the following, we allow for metric perturbations which change the determinant, i.e., $\delta\mathfrak{g}\ne0$.

A causal diamond possesses a conformal isometry generated by a conformal Killing vector, $\zeta^{\mu}$~\cite{Jacobson:2015}. In our coordinate system, $\zeta^{\mu}$ reads, up to an arbitrary normalisation constant, $C$,
\begin{equation}
\label{conformal killing}
\zeta=C\left(\left(l^2-t^2-r^2\right)\frac{\partial}{\partial t}-2tx^{i}\frac{\partial}{\partial x_{i}}\right),
\end{equation}
where $r=\sqrt{x_ix^{i}}$. The GLCD's boundary is a conformal Killing horizon with respect to $\zeta^{\mu}$. Its spatial cross-section at $t=0$ which corresponds to the boundary of the geodesic ball, $\partial\Sigma_0$, is a bifurcate surface and $\zeta^{\mu}$ vanishes there, as can be easily seen from equation~\eqref{conformal killing}. The surface gravity $\kappa$ corresponding to $\zeta^{\mu}$ obeys $\kappa=2lC$ at $\partial\Sigma_0$.

To provide a consistency test for our formalism, we first derive a relation between the volume of $\Sigma_0$ and the area of $\partial\Sigma_0$. The derivation proceeds similarly to that of the Smarr formula for black holes (see subsection~\ref{lambda}). We start by integrating the symplectic potential~\eqref{theta z} corresponding to $\zeta^{\mu}$ over the geodesic ball $\Sigma_0$
\begin{equation}
\int_{\Sigma_0}\theta^{\mu}\left[\delta_{\zeta}\right]\text{d}\Sigma_{\mu}=\int_{\Sigma_0}\Pi^{\mu}_{\zeta}\text{d}\Sigma_{\mu}+\int_{\partial\Sigma_0}\frac{1}{8\pi}\left(\sqrt{-\mathfrak{g}}/\omega\right)^{2/n}\tilde{\nabla}^{[\nu}\zeta^{\mu]}\text{d}\Sigma_{\mu\nu}.
\end{equation}
The symplectic potential $\theta^{\mu}\left[\delta_{\zeta}\right]$ can be worked out from equation~\eqref{theta z}. Since $\delta_{\zeta}\tilde{g}_{\mu\nu}=0$ we easily obtain $\theta^{\mu}\left[\delta_{\zeta}\right]=0$. Therefore, we have
\begin{equation}
\label{glcd smarr}
\int_{\Sigma_0}\Pi^{\mu}_{\zeta}\text{d}\Sigma_{\mu}+\int_{\partial\Sigma_0}\frac{1}{8\pi}\left(\sqrt{-\mathfrak{g}}/\omega\right)^{2/n}\tilde{\nabla}^{[\nu}\zeta^{\mu]}\text{d}\Sigma_{\mu\nu}=0.
\end{equation}
In flat (or any maximally symmetric) spacetime, it holds
\begin{equation}
\Pi^{\mu}_{\zeta}=-\frac{1}{8\pi}\frac{n}{n-2}k\kappa n^{\mu},
\end{equation}
where $k=\left(n-2\right)/l$ is the extrinsic curvature of the $(n-2)$-sphere $\partial\Sigma_0$ and $\kappa=2lC$ is the surface gravity corresponding to $\zeta^{\mu}$. Performing both integrals in equation~\eqref{glcd smarr} then leads to
\begin{equation}
\frac{1}{8\pi}\frac{n-1}{n-2}\kappa k\mathcal{V}-\frac{1}{8\pi}\kappa\mathcal{A}=0,
\end{equation}
where $\mathcal{A}$ and $\mathcal{V}$ denote the area of $\partial\Sigma_0$ and the volume of $\Sigma_0$, respectively. This result is essentially just the well-known flat space relation between the $(n-1)$-dimensional ball's volume and the area of its boundary, $\mathcal{A}=\left(n-1\right)\mathcal{V}/l$. Hence, its recovery serves to check the consistency of our formalism.

Next, we derive the first law of causal diamonds governing small perturbations of the metric from the flat spacetime to some other solution of the vacuum WTG equations of motion. Since $\delta_{\zeta}\tilde{g}_{\mu\nu}=0$, we can easily see that the symplectic current $\Omega^{\mu}\left[\delta_{\zeta},\delta\right]$ vanishes and, therefore, the perturbation of the Hamiltonian for the
evolution along $\zeta^{\mu}$ vanishes as well. Then, equation~\eqref{dH} directly yields the first law of causal diamonds
\begin{align}
\nonumber \delta 0=H_{\zeta}=&\int_{\mathcal{C}}\Omega^{\mu}\left[\delta_{\zeta},\delta\right]\text{d}\mathcal{C}_{\mu}=\frac{1}{8\pi}\kappa\delta\mathcal{A}-\frac{1}{8\pi}k\kappa\delta\mathcal{V}-\frac{1}{8\pi}\frac{n-2}{n}\kappa\int_{\partial\Sigma_0}\delta\ln\left(\sqrt{-\mathfrak{g}}/\omega\right)\text{d}^{n-2}x \\ &+\frac{1}{8\pi}\frac{n-1}{n}k\kappa\int_{\Sigma_0}\delta\ln\left(\sqrt{-\mathfrak{g}}/\omega\right)\text{d}^{n-1}x,
\end{align}
where $\delta\mathcal{A}$ and $\delta\mathcal{V}$ denote the perturbations of the area of $\partial\Sigma_{0}$ and the volume of $\Sigma_0$, respectively. In the unimodular gauge and for determinant preserving perturbations, $\delta\mathfrak{g}=0$, we reproduce the first law of causal diamonds valid in GR~\cite{Jacobson:2019}.

While the final form of the first law of causal diamonds in WTG is physically equivalent to the GR one, the term corresponding to the volume perturbation appears in a different way in both theories. In GR, the perturbation of the Hamiltonian is expressed entirely as a surface integral and the volume perturbation comes from a non-vanishing symplectic current, $\Omega^{\mu}\left[\delta_{\zeta},\delta\right]$~\cite{Jacobson:2019}. In other words, the perturbation of the Hamiltonian corresponding to the evolution along $\zeta^{\mu}$ does not vanish, $\delta H_{\zeta}\ne0$. While $\delta H_{\zeta}$ in GR always vanishes for true Killing vectors, it is in general nonzero for conformal Killing vectors. However, in WTG Weyl invariance of the symplectic current $\Omega^{\mu}\left[\delta_{\zeta},\delta\right]$ implies that it always vanishes for conformal Killing vectors and we have $\delta H_{\zeta}=0$. Instead of coming from $\Omega^{\mu}\left[\delta_{\zeta},\delta\right]$, the volume perturbation term instead appears in the first law from the volume integral,
\begin{equation}
\frac{n-2}{n}\int_{\mathcal{C}}\left(\frac{1}{16\pi}\tilde{\nabla}_{\nu}\tilde{\nabla}_{\rho}\zeta^{\rho}\delta\tilde{g}^{\mu\nu}+\theta^{\mu}\left[\delta\right]\tilde{\nabla}_{\nu}\zeta^{\nu}\right)\text{d}\mathcal{C}_{\mu},
\end{equation}
present in the perturbation of the Hamiltonian~\eqref{dH}.

\section{Discussion}
\label{discussion}

The principal outcome of the paper is an extension of the Noether charge formalism to WTG, a WTDiff invariant theory of gravity whose solutions are equivalent to those of GR. Moreover, we have employed it to obtain a statement of the first law of black hole mechanics, both in vacuum and in the presence of a perfect fluid, and an expression for WTG Wald entropy. The results turn out to be physically equivalent to those for GR. However, all the relevant physical quantities (total energy, total angular momentum, black hole entropy, etc.) are Weyl invariant and reduce to their form valid in GR only in the unimodular gauge. These findings precisely correspond to what one would expect, given that the equations of motion of WTG and GR coincide in the unimodular gauge. In this way, we provide a consistency check for the physical equivalence of both theories.

The equivalence of GR and WTG nevertheless breaks down in two respects. First, WTG allows coupling of gravity to certain locally non-conservative matter sources. Our results apply to such cases without any difficulties. Since breaking of local energy-momentum conservation is sometimes considered in unimodular and WTG cosmological models, our Noether charge formalism might serve to provide the appropriate conservation laws for them. Second, the cosmological term in WTG appears as an integration constant, its value is radiatively stable and, in general, it varies between solutions of the equations of motion. In this regard, we show that entropy of the de Sitter horizon in WTG and GR has the same value in the unimodular gauge. Furthermore, we derive a contribution of the varying cosmological constant to the first law of mechanics for a Schwarzschild-anti-de Sitter black hole. While such results have been previously obtained in GR, there they require an {\it ad hoc} assumption of a varying cosmological constant (for instance, by appealing to its interpretation as vacuum energy). In contrast, our approach has the advantage of naturally demanding a varying cosmological constant, already on the fully classical level.

While we derived several physically relevant results for WTG, the main purpose of our work is to introduce a new formalism useful for a systematic study of WTG. Some possible future applications of it include calculating Noether charges corresponding to the symmetries of the null infinity in asymptotically flat spacetimes, the Bondi-Metzner-Sachs group. Since calculations related to the Bondi-Metzner-Sachs group involve a conformal transformation of the metric, it might be interesting to see the impact of the Weyl invariance in this case. Another option lies in exploring the conserved quantities corresponding to dynamics of the background volume $n$-form and their impact on horizon thermodynamics (this will require a generalisation of the formalism to gauge field theories, along the lines of~\cite{Prabhu:2017}). Moreover, the Noether charge formalism for GR has been used to construct the solution phase space (see, e.g.~\cite{Adami:2021}). Obtaining the solution phase space for WTG in the same way might further clarify the (in)equivalence of both theories. We plan to address these issues in future works.

Originally, the Noether charge formalism was obtained for any local, Diff invariant theory of gravity, regardless of the form of its Lagrangian. In principle, it should be possible to similarly obtain a generalisation of the formalism we derived for WTG to any local, WTDiff invariant theory of gravity. We will address this in a forthcoming paper~\cite{Alonso:preparation}.

Lastly, an extra motivation for exploring WTG comes from thermodynamics of spacetime. It appears that from thermodynamic arguments there emerge gravitational dynamics equivalent to WTG. Within the presently developed formalism, we will be able to study this emergence more rigorously and possibly provide a conclusive proof of it. This would lend further support for the physical relevance of WTG as an alternative to GR.

\section*{Acknowledgments}

The authors want to acknowledge Gerardo Garc\'{i}a-Moreno for helpful discussions. AA-S is supported by the ERC Advanced Grant No. 740209. ML is supported by the Charles University Grant Agency project No. GAUK 297721. This work is also partially supported by the Spanish Government through Project. No. MICINN PID2020-118159GB-C44.


\begin{thebibliography}{99}
	
\bibitem{Wald:1990}
J. Lee and R.~M.~Wald,
``Local symmetries and constraints,''
J Math. Phys. {\bf 31} (1990) 725.

\bibitem{Wald:1993}
R.~M.~Wald,
``Black Hole Entropy is Noether Charge,''
Phys. Rev. D {\bf 48} (1993) 3427-3431 [gr-qc/9307038].

\bibitem{Wald:1994}
V. Iyer and R.~M.~Wald,
``Some Properties of Noether Charge and a Proposal for Dynamical Black Hole Entropy,''
Phys. Rev. D {\bf 50} (1994) 846-864 [gr-qc/9403028].

\bibitem{Iyer:1996}
V.~Iyer,
``Lagrangian perfect fluids and black hole mechanics,''
Phys. Rev. D {\bf 55} (1996) 3411-3426 [gr-qc/9610025].

\bibitem{Wald:2000}
R.~M.~Wald and A.~Zoupas,
``A General Definition of ``Conserved Quantities'' in General Relativity and Other Theories of Gravity,''
Phys. Rev. D {\bf 61} (2000) 084027 [gr-qc/9911095].

\bibitem{Alvarez:2006}
E.~\'{A}lvarez, D.~Blas, J.~Garriga and E. Verdaguer,
``Transverse Fierz-Pauli symmetry,''
Nucl. Phys. B {\bf 756} (2006) 148 [hep-th/0606019].

\bibitem{Barcelo:2014}
C.~Barcel\'{o}, R.~Carballo-Rubio and L.~J.~Garay,
``Unimodular gravity and general relativity from graviton self-interactions,''
Phys. Rev. D {\bf 89} (2014) 124019 [gr-qc/1401.2941].

\bibitem{Barcelo:2018}
C.~Barcel\'{o}, L.~J.~Garay and R.~Carballo-Rubio,
``Absence of cosmological constant problem in special relativistic field theory of gravity,''
Ann. Phys. {\bf 398} (2018) 9 [gr-qc/1406.7713].

\bibitem{Weinberg:1989}
S.~Weinberg,
``The cosmological constant problem,''
Rev. Mod. Phys. {\bf 61} (1989) 1.

\bibitem{Polchinski:2006}
J.~Polchinski,
``The Cosmological Constant and the String Landscape,''
[hep-th/0603249].

\bibitem{Burgess:2013}
C.~P.~Burgess,
``The Cosmological Constant Problem: Why it’s hard to get Dark Energy from Micro-physics,''
[hep-th/1309.4133].

\bibitem{Donoghue:2021}
J.~F.~Donoghue,
``The cosmological constant and the use of cutoffs,''
Phys. Rev. D {\bf 104} (2021) 045005 [gr-qc/2009.00728].

\bibitem{Padilla:2015}
A.~Padilla and I.~D.~Saltas,
``A note on classical and quantum unimodular gravity,''
Eur. Phys. J. C {\bf 75} (2015) [gr-qc/1409.3573].

\bibitem{Carballo:2015}
R.~Carballo-Rubio,
``Longitudinal diffeomorphisms obstruct the protection of vacuum energy,''
Phys. Rev. D {\bf 91} (2015) 124071 [gr-qc/1502.05278].

\bibitem{Jacobson:2019}
T.~Jacobson and M.~R.~Visser,
``Gravitational Thermodynamics of Causal Diamonds in (A)dS,''
SciPost Phys {\bf 7} (2019) 079 [hep-th/1812.01596].

\bibitem{Jacobson:2015}
T.~Jacobson,
``Entanglement Equilibrium and the Einstein Equation,''
Phys. Rev. Lett.  {\bf 116} (2016)	[gr-qc/1505.04753].

\bibitem{Bueno:2017}
P.~Bueno, V.~S.~Min, A.~J.~Speranza and M.~R.~Visser,
``Entanglement equilibrium for higher order gravity,''
Phys. Rev. D {\bf 95} (2017) 046003 [hep-th/1612.04374].

\bibitem{Svesko:2019}
A.~Svesko,
``Equilibrium to Einstein: Entanglement, Thermodynamics, and Gravity,''
Phys. Rev. D {\bf 99} (2019) 086006 [hep-th/1810.12236].

\bibitem{Alonso:2020a}
A.~Alonso-Serrano and M.~Li\v{s}ka,
``New Perspective on thermodynamics of spacetime: The emergence of unimodular gravity and the equivalence of entropies,''
Phys. Rev. D {\bf 102} (2020) 104056 [gr-qc/2008.04805].

\bibitem{Alonso:2021}
A.~Alonso-Serrano and M.~Li\v{s}ka,
``Thermodynamics of spacetime and unimodular gravity,''
[gr-qc/2112.06301].

\bibitem{preparation}
A.~Alonso-Serrano and M.~Li\v{s}ka,
in preparation.

\bibitem{MTW}
C.~W.~Misner, K.~S.~Thorne and J.~A.~Wheeler,
``Gravitation,''	
Princeton University Press, Princeton (2017).

\bibitem{Alvarez:2013a}
E.~\'{A}lvarez and M.~H.~Valea,
``Unimodular gravity with external sources,''
JCAP {\bf 2013} (2013) [hep-th/1209.6223].

\bibitem{Oda:2017}
I.~Oda,
``Classical Weyl Transverse Gravity,''
EPJ C {\bf 77} (2017) [hep-th/1610.05441].

\bibitem{Jirousek:2019}
P.~Jirousek and A. Vikman,
``New Weyl-invariant vector-tensor theory for the cosmological constant,''
JCAP {\bf 04} (2019) 004 [gr-qc/1811.09547].

\bibitem{Carballo:2020}
S.~Alexander and R.~Carballo-Rubio,
``Topological Features of the Quantum Vacuum,''
Phys. Rev. D {\bf 101} (2020) 024058 [gr-qc/1810.02159].

\bibitem{Josset:2017}
T.~Josset, A.~Perez and D.~Sudarsky,
``Dark Energy from Violation of Energy Conservation,''
Phys. Rev. Lett. {\bf 118} (2017) [gr-qc/1604.04183].

\bibitem{Perez:2018}
A.~Perez, D.~Sudarsky and J.~D.~Bjorken,
``A microscopic model for an emergent cosmological constant,''
Int. J. Mod. Phys. {\bf 27} (2018) [gr-qc/1804.07162].

\bibitem{Jackiw:2014}
R.~Jackiw and S.-Y.~Pi,
``Fake Conformal Symmetry in Conformal Cosmological Models,''
Phys. Rev. D  {\bf 91} (2015) 067501 [gr-qc/1407.8545].

\bibitem{Kubiznak:2015}
D.~Kubiznak and R.~B.~Mann,
``Black Hole Chemistry,''
Can. J. Phys. {\bf 93} (2015) 999 [gr-qc/1404.2126].

\bibitem{Hawking:1975}
S.~W.~Hawking,
``Particle Creation by Black Holes,''
Commun. Math. Phys. {\bf 43} (1975) 199-220.

\bibitem{Alvarez:2013b}
E.~\'{A}lvarez and M.~H.~Valea,
``No conformal anomaly in unimodular gravity,''
Phys. Rev. D  {\bf 87} (2013) 84054 [hep-th/1301.5130].

\bibitem{Visser:2003}
M.~Visser,
``Essential and inessential features of Hawking radiation,''
Int. J. Mod. Phys. D {\bf 12} (2003) 649 [hep-th/0106111].

\bibitem{Kastor:2009}
D.~Kastor, S.~Ray and J.~Traschen,
``Enthalpy and the Mechanics of AdS Black Holes,''
Class. Quant. Grav. {\bf 26} (2009) 195011 [hep-th/0904.2765].

\bibitem{Hawking:1996}
S.~W.~Hawking and G.~T.~Horowitz,
``The Gravitational Hamiltonian, Action, Entropy, and Surface Terms,''
Class. Quant. Grav. {\bf 13} (1996) 1487 [gr-qc/9501014].

\bibitem{Jager:2008}
S.~Jäger,
``Conserved quantities in asymptotically de Sitter spacetimes,''
Master thesis, University of Göttingen (2008).

\bibitem{Fiol:2010}
B.~Fiol and J.~Garriga,
``Semiclassical Unimodular Gravity,''
JCAP {\bf 2010} (2010) [hep-th/0809.1371].

\bibitem{Gibbons:1977}
G.~W~Gibbons and S.~W.~Hawking,
Cosmological event horizons, thermodynamics, and particle creation,
{\it Phys. Rev. D} {\bf 15} (1977) 2738.

\bibitem{Wald:1990b}
R.~M.~Wald,
``On identically closed forms locally constructed from a field,''
J Math. Phys.\ {\bf 31} (1990) 2378.

\bibitem{Bardeen:1973} 
J.~D.~Bardeen, B.~Carter and S.~W.~Hawking,
``The Four Laws of Black Hole Mechanics,''
Commun.\ Math.\ Phys. {\bf 31} (1973) 161.

\bibitem{Brown:1993}
D.~Brown,
``Action functionals for relativistic perfect fluids,''
Class. Quant. Grav.  {\bf 10} (1993) 1579-1606 [gr-qc/9304026].

\bibitem{Prabhu:2017}
K.~Prabhu,
``The First Law of Black Hole Mechanics for Fields with Internal Gauge Freedom,''
Class. Quant. Grav. {\bf 34} (2017) 035011 [gr-qc/1511.00388].

\bibitem{Adami:2021}
H.~Adami, M.~M.~Sheikh-Jabbari, V.~Taghiloo and H.~Yavartanoo,
``Null Surface Thermodynamics,''
[hep-th/2110.04224].

\bibitem{Alonso:preparation}
A.~Alonso-Serrano, L.~J.~Garay and M.~Li\v{s}ka,
``Noether charge formalism for Weyl invariant theories of gravity,''
in preparation.


\end{thebibliography}
\end{document}